\documentclass [12pt]{article}
\def\maj#1{\ifmmode\mbox{\usefont{U}{msb}{m}{n}#1}\else{\usefont{U}{msb}{m}{n}#1}\fi}
\def\v#1{\mathbf{#1}}
\makeatletter
\@addtoreset{equation}{section}
\makeatother
\renewcommand{\theequation}{\arabic{section}.\arabic{equation}}
\usepackage{graphics,epsfig,graphicx}
\usepackage{color}

\begin{document}

\title{\textbf{Exciton-exciton scattering :
\\ Composite boson versus elementary boson}}
\author{M. Combescot$^1$, O. Betbeder-Matibet$^1$ and R. Combescot$^2$
  \\ \small{\textit{$^1$ Institut des NanoSciences de Paris,}}\\
\small{\textit{Universit\'e Pierre et Marie Curie-Paris 6,
Universit\'{e} Denis Diderot-Paris 7, CNRS, UMR 7588,}}\\ 
\small{\textit{Campus Boucicaut, 140 rue de
Lourmel, 75015 Paris}}\\ \small{\textit{$^2$ Laboratoire de Physique
Statistique,}}\\ \small{\textit{Laboratoire de Physique de l'Ecole Normale
Sup\'{e}rieure,}}\\ \small{\textit{24 rue Lhomond, 75005 Paris}}}
\date{}
\maketitle

\begin{abstract}
This paper shows the necessity of introducing a new quantum object, the
``coboson'', to properly describe any composite particle, like the exciton,
which is made of two fermions. Although commonly dealed with as
elementary bosons, these composite bosons --- ``cobosons'' in short --- differ from them due to
their composite nature which makes the  handling of their many-body effects
quite different from the existing treatments valid for
elementary bosons. As a direct consequence of this composite nature,
there is no correct way to describe the interaction between cobosons
as a potential $V$. This is rather dramatic because, with the Hamiltonian not
written as $H=H_0+V$, all the usual approaches to many-body effects fail.
In particular, the standard form of
the Fermi golden rule, written in terms of $V$, cannot be used to obtain the transition rates of
two cobosons. To get them, we have had to construct an unconventional expression
for this Fermi golden rule in which $H$ only appears. Making use of this new
expression, we here give a detailed calculation of the time evolution
of two excitons. We compare the results of this exact approach with the
ones obtained by using an effective bosonic Hamiltonian in which the excitons are considered
as elementary bosons with effective scatterings between them, these scatterings resulting from an
elaborate mapping between the two-fermion space and the ideal boson space. We show that the
relation between the inverse lifetime and the sum of the transition rates for elementary bosons
differs from the one of composite bosons  by a factor of 1/2 ; so that it is impossible to find
effective scatterings between bosonic excitons giving these two physical quantities
correctly, whatever the mapping from composite bosons to elementary bosons is. The present paper
thus constitutes a strong mathematical proof that, in spite of a widely spread  belief, we cannot
forget the composite nature of these cobosons, even in the extremely low density limit of just two
excitons. This paper also shows the (unexpected) cancellation in the Born approximation of the
two-exciton transition rate for a finite value of the momentum transfer.
\end{abstract}

\vspace{0.5cm}


\newpage

\section{Introduction}

Excitons are known to be composite particles made of one electron and
one hole. They of course interact through the electron-electron,
hole-hole and electron-hole Coulomb potentials. They also interact in a
quite subtle manner, through Pauli exclusion between the fermions from which
they are made. The purpose of this paper is to study
how this Pauli exclusion enters the exciton transition rate and lifetime.

The major difficulty induced by the composite nature of the
excitons is the impossibility to identify an interaction potential
\emph{between} excitons, even for the Coulomb contribution.
While the electron-electron and hole-hole Coulomb potentials are
unambiguously parts of the interaction between two excitons, such an identification is ambiguous
for the electron-hole parts. Indeed, while $V_{eh'}=
-e^2/|\v r_e-\v r_{h'}|$ is part of the interaction between excitons made of $(e,h)$ and $(e',h')$,
the same
$V_{eh'}$ is clearly \emph{not} part of the interaction between excitons if these excitons are made
of
$(e,h')$ and
$(e',h)$. Since electrons and holes are indistinguishable, there is no
way to know how these two excitons are made, so that there is no way to
write the part of the Coulomb interaction between two excitons properly.

In spite of this obvious problem, various procedures [1,2] have been
proposed to replace the semiconductor Hamiltonian written in terms of
electrons and holes by an effective Hamiltonian written in terms of
excitons considered as elementary bosons, with an effective exciton-exciton
potential between them. Even if the bosonization procedures may appear as
rather sophisticated [3], it is clear that some uncontrolled manipulations have
to be done in the mapping of the two-fermion subspace into the ideal boson subspace, in order to
transform the exact electron-electron Coulomb potential, written in terms of electron operators as
$a^\dag a^\dag a\,a$, into a part of an exciton-exciton potential, which,
in terms of exciton operators, reads as $B^\dag B^\dag B\,B$, these
exciton operators being made of electron-hole pairs, \emph{i}.\
\emph{e}., $B^\dag$ being linear combinations of $a^\dag b^\dag$. In a
previous work [4], we have already shown that there is no way to find
prefactors for these
$B^\dag B^\dag B\,B$ terms which could produce the correct
correlations between two excitons at any order in the exciton-exciton
interaction. We here show somewhat in details that there is no way to
find $B^\dag B^\dag B\,B$ prefactors which would produce
\emph{both} the exciton-exciton transition rate \emph{and} the lifetime of
an  exciton state correctly, even in the case of just two excitons : 
If we know the correct value of one of these two physical
quantities, for example the transition rate, we can possibly adjust the
$B^\dag B^\dag BB$ prefactors in the effective  bosonic Hamiltonian to
recover the correct transition rate. However, there is no way to be sure that
the {\it same} prefactors would give other
physical quantities correctly, a factor of 2 being actually missed in the 
lifetime, as previously reported in ref.\ [5]. 

This major failure, which
puts the concept of effective bosonic Hamiltonian to describe interacting excitons in a very bad
shape, should push the very large amount of physicists using such effective Hamiltonians [6-14],
to reconsider their works in the light of the many-body theory for ``cobosons'' --- a
contraction for composite bosons --- that we have recently constructed
[15] and which is free from any bosonization.

The fact that a trustworthy exciton-exciton potential does not exist, has
dramatic consequences on the possible treatment of many-body effects involving
excitons. Indeed, all known approaches to many-body effects [16,17]
are based on rewriting the Hamiltonian as
$H=H_0+V$, with $V$ being the interaction potential responsible for the
many-body effects, and treating $V$ as a perturbation, possibly at
infinite order in case of singularities. Without the availability of
such a  potential $V$, a novel many-body procedure, which does not rely
on a would-be
$V$, has thus to be constructed from scratch in order to derive many-body
effects between excitons. This is the purpose of the theory we are
presently developing.

In order to put the difficulty associated with the composite nature of
the excitons on a proper formal basis, it is of importance to realize
that all failures in the previous approaches to the exciton 
many-body physics can be traced back to the difficulty of properly
handling Pauli exclusion, which prevents these excitons  from
being exact bosons. Let us introduce the exciton creation operators
$B_i^\dag$ as being such that the $B_i^\dag|v\rangle$'s are the exact one-electron-hole-pair
eigenstates of the semiconductor Hamiltonian,
$$(H-E_i)B_i^\dag|v\rangle=0,$$ 
$|v\rangle$ being the vacuum state. These operators can be expanded on the 
free-electron-hole-pair operators as
\begin{equation}
B_i^\dag = \sum_{\v k_e, \v k_h}\langle \v k_e,\v k_h|i\rangle
a_{\v k_e}^\dag\,b_{\v k_h}^\dag\ ,
\end{equation}
where $a_{\v k_e}^\dag$ and $b_{\v k_h}^\dag$ are the creation operators for
free electrons and free holes, with momentum $\v k_e$ and $\v
k_h$, respectively.
$\langle
\v k_e,\v k_h|i\rangle$ is the $i$ exciton wave
function in momentum space, namely $\langle \v k_e,\v
k_h|i\rangle=\delta_{\v k_e+\v k_h,\v Q_i}\,
\langle \alpha_h\v k_e-\alpha_e\v k_h|\nu_i\rangle$, where $\v Q_i$ is the $i$ exciton
center-of-mass momentum and
$\langle\v k|\nu_i\rangle$ is the relative motion wave
function of the
$i$ exciton in $\v k$ space, with $\alpha_e=1-\alpha_h=m_e/(m_e+m_h)$.
Using eq.\ (1.1), it is straightforward to show that, while $[B_i,B_j]=0$ as usual for
bosons, the commutator
$\left[B_i,B_j^\dag\right]$ differs from $\delta_{ij}$, which would be
its value if the excitons were elementary bosons.

In order to set up on a precise ground the formalism associated to the
fact that excitons differ from elementary bosons, we have been led
to introduce [18,19] the set of Pauli parameters $\lambda\left(^{n\ j}_{m\
i}\right)$ defined as
\begin{equation}
\left[D_{mi},B_j^\dag\right]=\sum_n\left[\lambda\left(^{n\ j}_{m\
i}\right)+\lambda\left(^{m\ j}_{n\
i}\right)\right]\,B_n^\dag\ ,
\end{equation}
where the $D_{mi}$'s are ``deviation-from-boson operators'' defined as
\begin{equation}
\left[B_m,B_i^\dag\right]=\delta_{mi}-D_{mi}\ .
\end{equation}
The physical understanding of these $\lambda\left(^{n\ j}_{m\
i}\right)$ parameters and their link with the exciton composite nature
become transparent once their expressions in real space is
given : As rederived in appendix A, these parameters read
\begin{eqnarray}
\lambda\left(^{n\ j}_{m\ i}\right)
&=& \int d\v r_e\,d\v r_{e'}\, d\v
r_h\,d\v r_{h'}\,\phi_m^\ast(\v
r_e,\v r_{h'})\,\phi_n^\ast(\v r_{e'},\v r_h)\,\phi_i(\v r_e,\v r_h)
\,\phi_j(\v r_{e'},\v r_{h'})\nonumber \\
&=& \lambda\left(^{m\ i}_{n\
j}\right)=
\lambda\left(^{j\ n}_{i\
m}\right)^\ast\ ,
\end{eqnarray}
where the $i$ exciton wave function in real space is $\phi_i(\v r_e,\v
r_h)=\langle
\v r_e,\v r_h|i\rangle\linebreak= \langle\alpha_e\v r_e+\alpha_h\v
r_h|\v Q_i\rangle\,\langle\v r_e-\v r_h|
\nu_i\rangle$, with $\langle\v R|\v Q\rangle=e^{i\v Q.\v R}/L^{D/2}$, for 
sample size $L$ and space dimension $D$. The Pauli parameter
$\lambda\left(^{n\ j}_{m\ i}\right)$ corresponds to a hole exchange
when forming the ``out'' excitons
$(m,n)$ from the ``in'' excitons $(i,j)$ (see Fig.1a).
These exchange parameters are not exactly exchange
scatterings in the sense that they are dimensionless.

One important property of the $\lambda\left(^{n\ j}_{m\ i}\right)$
parameters is the fact that if we cross the holes of two excitons
twice, we come back to the original situation. As shown in
appendix A, we indeed have
\begin{equation}
\sum_{mn}\lambda\left(^{q\ n}_{p\
m}\right)\,\lambda\left(^{n\ j}_{m\
i}\right)=\delta_{pi}\,\delta_{qj}\ .
\end{equation}

Another transparent link between $\lambda\left(^{n\ j}_{m\ i}\right)$
and the possibility to exchange the carriers in forming two
excitons is the fact that we can rewrite any $B_i^\dag B_j^\dag$ in
terms of all the other
$B_m^\dag B_n^\dag$'s according to
\begin{equation}
B_i^\dag B_j^\dag=-\sum_{mn}\lambda\left(^{n\ j}_{m\
i}\right)\,B_m^\dag B_n^\dag\ .
\end{equation}
As shown in appendix A, this equation is obtained by writing
the
$i$ exciton in terms of $a_{\v k_{e_i}}^\dag b_{\v k_{h_i}}^\dag$ and by
forming the $m$ exciton out of $a_{\v k_{e_i}}^\dag b_{\v k_{h_j}}^\dag$.

We expect the physics associated to the non-purely bosonic nature of the
excitons to appear through exchange processes of various kinds, which
are all going to be expressed in terms of these $\lambda\left(^{n\ j}_{m\
i}\right)$'s. The pure-bosonic exciton approximation used in
the effective bosonic Hamiltonian corresponds to take all these
$\lambda\left(^{n\ j}_{m\ i}\right)$'s equal to zero, \emph{after} having somehow
cooked them with Coulomb processes, once and for all, to produce an
exciton-exciton scattering ``dressed by exchange''.

In addition to non-bosonic behavior, the fact that the excitons are
made of indistinguishable carriers makes the Coulomb interaction
\emph{between} excitons quite tricky to define properly, as
discussed above. We however need to identify such a quantity
in a formal way, if we want to set up a procedure for handling many-body
effects between excitons, since of course they are going to contain a
certain amount of Coulomb processes. By writing  [18,19]
\begin{equation}
\left[H,B_i^\dag\right]=E_i\ B_i^\dag+V_i^\dag\ ,
\end{equation}
\begin{equation}
\left[V_i^\dag,B_j^\dag\right]=\sum_{mn}\xi^\mathrm{dir}\left(^{n\
j}_{m\ i}\right)\, B_m^\dag B_n^\dag\ ,
\end{equation}
we in fact generate the set of Coulomb scatterings we
want. Indeed, as rederived in appendix B, $\xi^\mathrm{dir}
\left(^{n\ j}_{m\ i}\right)$ reads as
\begin{eqnarray}
\xi^\mathrm{dir}\left(^{n\ j}_{m\
i}\right)=\int
d\v r_e\ d\v r_{e'}\,d\v r_h\,d\v r_{h'}\,
\phi_m^\ast(\v r_e,\v r_h)\,\phi_n^\ast(\v
r_{e'},\v r_{h'})
\left[V_{ee'}+V_{hh'}-V_{eh'}-V_{he'}\right]\nonumber\\ \times
\phi_i(\v r_e,\v
r_h)\,\phi_j(\v r_{e'},\v r_{h'})\nonumber\\
=\xi^\mathrm{dir}\left(^{m\ i}_{n\
j}\right)=\xi^\mathrm{dir}\left(^{j\ n}_{i\
m}\right)^\ast\ ,\hspace{7.3cm}
\end{eqnarray}
with $V_{cd'}=e^2/|\v r_c-\v r_{d'}|$.
We see that, in this $\xi^\mathrm{dir}\left(^{n\ j}_{m\
i}\right)$, the ``in'' exciton $i$ and ``out''
exciton $m$ are made with the same pairs $(e,h)$ and similarly for the excitons $(j,n)$ ;
so that, in $\xi^\mathrm{dir}\left(^{n\ j}_{m\
i}\right)$, the electron-hole
Coulomb interaction
$(V_{eh'}+V_{e'h})$ is unambiguously a Coulomb interaction
\emph{between} both the ``in'' excitons $(i,j)$ and the ``out'' excitons $(m,n)$ (see Fig.1b). Let
us add that, while the exchange parameters $\lambda\left(^{n\ j}_{m\
i}\right)$ are dimensionless, these
$\xi^\mathrm{dir}\left(^{n\ j}_{m\
i}\right)$'s are energy-like quantities.

Using these $\xi^\mathrm{dir}\left(^{n\ j}_{m\
i}\right)$'s and
$\lambda\left(^{n\ j}_{m\
i}\right)$'s, it is possible to derive the many-body
physics of excitons in an exact way. In particular, we can construct the two
exchange Coulomb scatterings which exist between two excitons. As rederived in
appendix C, the one in which the Coulomb interaction takes place between
the ``in'' excitons $(i,j)$ only reads
\begin{eqnarray}
\xi^\mathrm{in}\left(^{n\
j}_{m\ i}\right)&=&\sum_{rs}\lambda\left(^{n\
s}_{m\ r}\right)\,\xi^\mathrm{dir}\left(^{s\
j}_{r\ i}\right)
\nonumber \\ &=&
\int d\v r_e\ d\v r_{e'}\,d\v r_h\,d\v r_{h'}\,
\phi_m^\ast(\v r_e,\v r_{h'})\,\phi_n^\ast(\v
r_{e'},\v r_h)
\nonumber
\\ &\times &\left[V_{ee'}+V_{hh'}-V_{eh'}-V_{he'}\right]\phi_i(\v
r_e,\v r_h)\,
\phi_j(\v r_{e'},\v r_{h'})\nonumber
\\ &=& \xi^\mathrm{in}\left(^{m\
i}_{n\ j}\right)\ .
\end{eqnarray}
In this Coulomb scattering,
the electron-hole part $(V_{eh'}+V_{e'h})$ is between the ``in''
excitons $(i,j)$ but ``inside'' the ``out'' excitons $(m,n)$ (see Fig.1c). In a
similar way,
\begin{eqnarray}
\xi^\mathrm{out}\left(^{n\
j}_{m\ i}\right)&=&\sum_{rs}\xi^\mathrm{dir}\left(^{n\
s}_{m\ r}\right)\,
\lambda\left(^{s\
j}_{r\ i}\right)
\nonumber \\ &=&
\int d\v r_e\ d\v r_{e'}\,d\v r_h\,d\v r_{h'}\,
\phi_m^\ast(\v r_e,\v r_h)\,\phi_n^\ast(\v
r_{e'},\v r_{h'})
\nonumber
\\ &\times &\left[V_{ee'}+V_{hh'}-V_{eh'}-V_{he'}\right]\phi_i(\v
r_e,\v r_{h'})\,
\phi_j(\v r_{e'},\v r_h)\nonumber
\\ &=& \xi^\mathrm{out}\left(^{m\
i}_{n\ j}\right)=\left[\xi^\mathrm{in}\left(^{j\
n}_{i\ m}\right)\right]^\ast\ ,
\end{eqnarray}
contains all Coulomb interactions between the ``out'' excitons $(m,n)$, its
electron-hole part being inside the ``in'' excitons $(i,j)$ (see Fig.1d).

The exchange parameters $\lambda\left(^{n\ j}_{m\ i}\right)$ being
dimensionless, it is possible to build 
energy-like scatterings out of them, through
\begin{equation}
\mathcal{E}\left(^{n\
j}_{m\ i}\right)=(E_m+E_n-E_i-E_j)\,\lambda\left(^{n\
j}_{m\ i}\right)\ .
\end{equation}
Note that, as the exciton
energy contains the band gap, scatterings having the sum of the ``in'' and ``out''
energies, instead of the difference, would depend on the band gap, which is physically unacceptable
for many-body effects coming from carrier interactions. Actually, such scatterings never appear.
It turns out that
$\mathcal{E}\left(^{n\ j}_{m\ i}\right)$ defined in eq.\ (1.12) is not an
independent scattering but reads in terms of the two exchange
Coulomb scatterings defined in eqs.\ (1.10-11) as (see appendix D)
\begin{equation}
\mathcal{E}\left(^{n\
j}_{m\ i}\right)=\xi^\mathrm{in}\left(^{n\
j}_{m\ i}\right)-\xi^\mathrm{out}\left(^{n\
j}_{m\ i}\right)\ .
\end{equation}

From the above discussion, we see that four scatterings between two
excitons are energy-like quantities, namely
$\xi^\mathrm{dir}\left(^{n\
j}_{m\ i}\right)$, $\xi^\mathrm{in}\left(^{n\
j}_{m\ i}\right)$,
$\xi^\mathrm{out}\left(^{n\
j}_{m\ i}\right)$ and $\mathcal{E}\left(^{n\
j}_{m\ i}\right)$. It is reasonable to expect
the exciton-exciton transition rates to read in terms of a linear
combination of these four scatterings. The purpose of the present work
is to determine this linear combination by using a full-proof procedure.

In order to do it, the first difficulty is to identify the proper way to determine these
transition rates. If an exciton-exciton potential were to exist, the exciton-exciton transition
rate would result from the Fermi golden rule written in terms of this potential. As such
a potential does not exist, it is necessary to construct a formal equivalent of
this Fermi golden rule in which $H$ only enters,
\emph{i}.\
\emph{e}., in which $H$ is not split as
$H_0+V$. This formal equivalent, already given in ref.\ [5], is rederived in
the section 2 of this paper, somewhat in details, for completeness.

In order to calculate these transition rates, we also need to identify the relevant ``in'' and
``out'' states of the transition. For that, we first note that an 
$N$-electron-hole-pair state can always be written in terms of free pair states, \linebreak$a_{\v
k_{e_1}}^\dag b_{\v k_{h_1}} ^\dag\cdots a_{\v k_{e_N}}^\dag b_{\v k_{h_N}}^\dag|v\rangle$, the
representation on this free pair basis being unique. However, as
electron-hole pairs are highly correlated into excitons
when their density in Bohr radius unit is small compared to 1, the
representation of physical  relevance for
$N$-pair states in the low density limit, is for sure not the one in terms
of free pair states but the one in terms of
exciton states, namely
$B_{i_1}^\dag\cdots B_{i_N}^\dag|v\rangle$.
This physically relevant representation however is
mathematically unpleasant because, due to eq.\ (1.6), it is not
unique, the $N$-exciton basis being overcomplete. As shown in details below, it is
actually possible to deal with this unpleasant feature and to calculate
the time evolution of any of these
$N$-exciton states, using the expression of the Fermi golden rule given in section 2. 

Among these $N$-exciton states, the state
$(B_0^\dag)^N|v\rangle$, with all the excitons in the same state 0, is
particularly simple and of possible physical interest. Indeed, while it is
\emph{not} the exact ground state of $N$ excitons --- otherwise it would not evolve with time,
--- it is close to it. Moreover, this state is the one coupled to
$N$ photons tuned to the ground state exciton, so that it plays an
important role in all semiconductor optical nonlinear effects. 

Although the state
$(B_0^\dag)^N|v\rangle$ may appear as particularly simple, let us stress that the precise
calculation of its time evolution already contains three major difficulties : (i) The first one is
to correctly determine  the time evolution of composite bosons taking into
account all the carrier exchanges which can take place between them. (ii) The
second one comes from the difficulty of handling many-body effects
between a large number of these composite particles. (iii) The third one comes
from the fact that excitons have a spin degree of freedom. As the
exchange processes mix the electrons and holes of two excitons, they also
mix their spins. They, in particular, transform two bright excitons with
opposite spins,
$(\pm 1)$, into two dark excitons with opposite spins $(\pm 2)$. If we
take into account these spin degrees of freedom, the exchange parameter
$\lambda\left(^{n\ j}_{m\ i}\right)$, defined
through eq.\ (1.2), becomes a
$8^2\times 8^2$ matrix, each bulk exciton having $2\times 4$
spin degrees of freedom. The situation is somewhat better for narrow
quantum wells since the light hole band with spins $(\pm 1/2)$ is
well separated from the heavy hole band with spins $(\pm 3/2)$, so
that we can forget it. However, the
$\lambda\left(^{n\ j}_{m\ i}\right)$ parameter is still
a 16$\times$16 matrix.

In order to reach a deep understanding of the tricky physics  involved
in these exciton scatterings, we find appropriate to divide this
work into three parts.

\underline{The present paper I} mainly deals with the
composite character of the excitons by considering the time
evolution of \emph{two excitons without spin degree of freedom}. This
physically corresponds to have two electrons with same spin and two holes
with same spin, as possibly produced by the absorption of two circularly
polarized photons in a quantum well. 

In section 3, we calculate its time
evolution using the formalism of the effective bosonic Hamiltonian, as many
physicists commonly think about excitons in this way. 

In section 4, we
calculate the time evolution of the state
$(B_0^\dag)^2|v\rangle$ made of two identical composite excitons. From
it, we derive the lifetime of this state as well as the transition rate
towards another two-exciton state $B_i^\dag B_j^\dag|v\rangle$, using
the formal equivalent of the Fermi golden rule rederived in the section 2 of this paper.

In section 5, we qualitatively discuss the various results obtained by using
the effective Hamiltonian in which the excitons are replaced by
elementary bosons, with similar quantities obtained for composite
excitons. 

In section 6, we quantitatively calculate the
various elementary scatterings that our composite exciton formalism
introduces, namely
$\xi^\mathrm{dir}$,
$\xi^\mathrm{in}$,
$\xi^\mathrm{out}$ and $\mathcal{E}$, in the case of 2D ground
state excitons, when the two ``in'' excitons have the same momentum. We
also calculate the transition rate of this two-composite-exciton state, as a function
of momentum transfer $Q$, and we compare it to its value when excitons are replaced by elementary
bosons. We see that the discrepancy is quite large, except for very small momentum transfer or in
the limit of infinitely heavy holes. This strongly questions the impressive fits of experimental
results obtained by using the standard exciton-exciton scatterings dressed by exchange, since in
real experiments, the hole mass is never that large, to make all exchange scatterings equal. We
also see that the exciton-exciton transition rate does not monotically decrease with increasing
momentum transfer, but cancels for a finite value of the momentum transfer $Q$. This cancellation
may suggest that, in order to get a physically relevant transition rate at large momentum
transfer, it might be necessary to go beyond the Fermi golden rule,
\emph{i.\ e.}, the Born approximation. However, it might also be possible that this cancellation
survives beyond Born approximation.

The discrepancy between the results obtained for elementary and
composite bosons is not fortuitous but has a quite deep origin. Indeed, the bosonic
approach has to fail in an irretrievable way because of a mathematical
reason. Bosonic exciton states form an orthogonal basis for two-pair
states. On the opposite, composite exciton states form an overcomplete
set, due to the composite nature of the particles. This overcompleteness
is directly linked to the appearance of an additional factor 1/2 between
the inverse lifetime and the sum of transition rates, which comes from the
difference which exists in the closure relations of composite and
elementary bosons, as explicitly shown in ref.\ [20]. The existence of
this factor 1/2 in fact shows that
\emph{it is impossible to build a set of effective scatterings
which would produce both, the lifetime and the transition rate towards
another exciton state, correctly. This
actually constitutes a strong mathematical proof that we cannot forget the
composite nature of the particles, even in the extremely low density limit
of just two excitons.}

Since the calculations with composite excitons we present here, use
quantities we have introduced in various previous works
dealing with what we first called ``commutation technique'', we have
found useful to rederive the important relations between these
quantities in a self-contained appendix with coherent notations, some of
these derivations being actually simpler than the ones we first
gave.

\underline{In paper II}, we will still consider two excitons only, but we will take
into account the spin degrees of freedom of these excitons. From the
interplay between direct and exchange scatterings, it is possible to deduce a
set of interesting polarization effects between the photons which create
the initial state and the bright and dark states produced by its time
evolution. In this paper II, we will consider all possible
polarizations for quantum wells only. In the case of bulk samples, it is
more appropriate to speak in terms of polaritons, instead of
excitons. The physics of interacting polaritons is \emph{a priori} very
similar to the one of interacting excitons, the additional photon part
of the polariton of course being an elementary boson. We are soon going to propose a new approach
to interacting polaritons. This will allow us to
derive the various subtle polarization effects which come from the fact that exchange
processes between the exciton parts of the polaritons are described by exchange parameters
$\lambda\left(^{n\ j}_{m\ i}\right)$ which
now are 64$\times$64 matrices.

\underline{Paper III} will deal with $N$-exciton states. The change from 2
to $N$ is not small; it induces some very substantial
difficulties. One important --- but quite tricky --- aspect of the $N$-exciton physics is the fact
that, unlike for $N$ electrons, 
many-body effects with $N$ excitons coming from exciton-exciton interaction are not associated to
an order in Coulomb interaction, but to an order in the dimensionless parameter associated to
density, namely,
$$\eta=N(a_X/L)^D,$$ 
with
$a_X$ being the exciton Bohr radius, $L$ the sample size and $D$ the
space dimension. The factor
$N$ entering this $\eta$ parameter can appear in calculations dealing with $N$ excitons in very
many different ways. A careful counting of these
$N$'s --- and the exact cancellation of overextensive terms in
$N^p(a_X/L)^{qD}$ with $p>q$ --- turns out to be rather tricky. However,
this is only in this last paper on the time evolution of
$N$-exciton states that we will really face the
many-body physics of composite excitons in its full
complexity. A short report on the scatterings between $N$ excitons,
without spin degree of freedom, has been published in ref.\ [5]. Its key
result is the fact that the additional factor $1/2$ between the
inverse lifetime and the sum of transition rates here shown for two excitons
also exists for
$N$ excitons. In a more recent work [20], we have established the link
between this additional $1/2$ factor and the overcompleteness of the
basis made with composite exciton states, by showing, in details,
how this factor $1/2$ appears for $N=2$ and $N=3$ excitons, using the
difference which exists in the closure relations of elementary and composite exciton
states.

The precise comparison between composite excitons and pure-bosonic
excitons done in the present paper, leads us to think that 
the present ``coboson'' formalism must be of
interest, not only for interacting excitons, but also for many
other composite bosons : This coboson formalism, free from bosonization,  might reveal unexpected
physical effects or, at least, lead to a deeper understanding of the
presently known physics. A first idea is to
reinvestigate interacting hydrogen atoms, or to
reconsider the physics of ``ultracold atoms'', which is a domain of
very high current interest. We can also think of using it in the on-going extensive studies of
interactions between positronium atoms [21] : These atoms seem to be very good candidates to reveal
the importance of fermion exchanges as they are formed with fermions having equal masses.

An easy switch from excitons to general cobosons is made by noting that the many-body physics of
any other coboson is expected to depend on exchange parameters $\lambda\left(^{n\
j}_{m\ i}\right)$ and direct scatterings
$\xi^\mathrm{dir}\left(^{n\ j}_{m\ i}\right)$, very similar to the ones
defined in eqs.\ (1.4) and (1.9), where $(\v r_e,\v r_h)$ represent the
spatial coordinates of the fermion pair at hand, $\phi_i(\v r_e,\v r_h)$ being
the wave function of the one-pair eigenstate of the system Hamiltonian and $V_{ee'}$,
$V_{hh'}$,
$V_{eh}$ the potentials between identical and different fermions. 

With respect to the possible representation of the coboson many-body physics, we can note that the
many-body theories at hand up to now [15,16] were designed to deal with interacting elementary
particles, fermions or bosons ; so that the Feynman diagrams which visualize the underlying 
perturbative theory, are rather easy to draw, due to the well
defined interaction potentials which exist between these elementary quantum particles.
For cobosons, we have had to construct, not only a new many-body theory, 
but also a fully new diagrammatic representation [22]. We have
called these new diagrams ``Shiva diagrams'', in reference to the
multiarm hindu god Shiva, as they have a multiarm structure. These diagrams
make appearing the elementary scatterings between two cobosons
$\xi^\mathrm{dir}\left(^{n\ j}_{m\ i}\right)$ and $\lambda\left(^{n\
j}_{m\ i}\right)$. It is of importance to stress that, since an
interaction potential between cobosons does not exist, the
corresponding diagrams are not a simple visualization of a perturbative theory.

\section{The Fermi golden rule in terms of H only}

Let us consider that, at initial time $t=0$, the system is in a
normalized initial state $|\psi_0\rangle$ which is not eigenstate of the
system Hamiltonian $H$, otherwise it would not evolve with time. At time
$t>0$, this state reads 
\begin{equation}
|\psi_t\rangle=e^{-i\hat{H}t}|\psi_0\rangle\equiv|\psi_0\rangle+|\tilde{\psi}
_t\rangle\ .
\end{equation}
By definition, $|\tilde{\psi}_t\rangle$ is the state change due to the time
evolution, while
$\hat{H}=H-\langle H\rangle_0$, with
$\langle H\rangle_0=
\langle\psi_0|H|\psi_0\rangle$ being the expectation value of the
Hamiltonian in the initial state (we take $\hbar=1$ throughout the
paper) : By writing $\hat{H}$ instead of $H$ in eq.\ (2.1), we have
in $|\psi_t\rangle$ introduced an irrelevant constant phase factor $e^{i\langle H\rangle_0t}$ for
convenience. 

We then note that
\begin{equation}
\hat{H}|\psi_0\rangle\equiv P_{\perp}H|\psi_0\rangle\ ,
\end{equation}
where $P_{\perp}=1-|\psi_0\rangle\langle\psi_0|$ is the projector over
the subspace perpendicular to $|\psi_0\rangle$, i.e., $P_{\perp}|\psi_0\rangle=0$. It is then easy
to  check that the state change $|\tilde{\psi}_t\rangle$, which physically comes
from scatterings in the initial state, can be rewritten as
\begin{equation}
|\tilde{\psi}_t\rangle=(e^{-i\hat{H}t}-1)|\psi_0\rangle=
F_t(\hat{H})\,P_\perp\,H\,|\psi_0\rangle\ ,
\end{equation}
where we have set
\begin{eqnarray}
F_t(E)&=&\frac{e^{-iEt}-1}{E}=-2i\pi\,e^{-iEt/2}\,\delta_t(E)\nonumber\\
\delta_t(E)&=&\frac{\sin(Et/2)}{\pi E}\ ,
\end{eqnarray}
$\delta_t(E)$ being a peaked function of width $2/t$, which reduces
to the usual Dirac $\delta$ function in the limit of infinitely
large time $t$.

From the definition of the lifetime
$\tau_0$ of the $|\psi_0\rangle$ state, namely, $|\langle\psi_0|\psi_t\rangle|^2=e^{-t/\tau_0}$,
we have, since $\langle\psi_t|\psi_t\rangle$ stays equal to 1,
\begin{eqnarray}
\frac{t}{\tau_0}&\simeq& 1-|\langle\psi_0|\psi_t\rangle|^2\nonumber\\
&=&\langle\psi_t
|\psi_t\rangle-|\langle\psi_0|\psi_t\rangle|^2=\langle\psi_t|P_{\perp}|\psi_t\rangle\ .
\end{eqnarray}
By using the state change
$|\tilde{\psi}_t\rangle$ introduced in eq.\ (2.1), and the fact that $P_{\perp}|\psi_t\rangle=
P_{\perp}|\tilde{\psi}_t\rangle$, eq.\ (2.5) can be rewritten as
\begin{equation}
\frac{t}{\tau_0}=\langle\tilde{\psi}_t|\tilde{\psi}_t\rangle-
|\langle\psi_0|\tilde{\psi}_t\rangle|^2\ .
\end{equation}

It is straightforward to check that the above eqs.\ (2.3) and (2.6) give
the well known result in the usual case, \emph{i}.\
\emph{e}., when $H=H_0+V$, with $|\psi_0\rangle\equiv |0\rangle$ and
$H_0|n\rangle=\mathcal{E}_n|n
\rangle$. Indeed, the state change at first order in $V$ is obtained
by replacing, in eq.\ (2.3), $F_t(\hat{H})$ by $F_t(\hat{H}_0)$, with
$\hat{H}_0=H_0-\mathcal{E}_0$, while $P_{\perp}H|0\rangle=\sum_{n\neq 0}
V_{n0}|n\rangle$, so that
\begin{eqnarray}
|\tilde{\psi}_t\rangle&=&F_t(\hat{H}_0)\sum_{n\neq
0}V_{n0}\,|n\rangle+0(V^2)\nonumber\\
&=&\sum_{n\neq 0}V_{n0}\,
F_t(\mathcal{E}_n-\mathcal{E}_0)\,|n\rangle+0(V^2)\ .
\end{eqnarray}
The transition rate from $|0\rangle$ to $|n\neq 0\rangle$ follows
from
\begin{equation}
\frac{t}{T_n}= |\langle n|\psi_t\rangle|^2=
|\langle n|\tilde{\psi}_t\rangle|^2\simeq 2\pi\,t
|V_{n0}|^2\,\delta_t(\mathcal{E}_n-\mathcal{E}_0)\ ,
\end{equation}
since we do have
\begin{equation}
|F_t(E)|^2\simeq|F_t(0)F_t(E)|\simeq|F_t(E)|t
=2\pi\, t\,\delta_t(E)\ .
\end{equation}
If we now turn to the lifetime of the $|0\rangle$ state, we find from
eq.\ (2.7) that $\langle 0|\tilde{\psi}_t\rangle=0(V^2)$, so that
$t/\tau_0$ to second order in $V$ reduces to $\langle\tilde{\psi}_t|
\tilde{\psi}_t\rangle$. Using eq.\ (2.7) and eq.\ (2.9), we thus recover
the well known result,
\begin{equation}
\frac{1}{\tau_0}\simeq 2\pi\,\sum_{n\neq 0}|V_{n0}|^2\,
\delta_t(\mathcal{E}_n-\mathcal{E}_0)\ .
\end{equation}
We can also get this equation by noting that, as the
$|n\rangle$ states are eigenstates of $H_0$, they form an orthonormal
basis, so that
\begin{equation}
1=\langle\psi_t|\psi_t\rangle=\langle\psi_t|\sum_n|n\rangle\langle
n||\psi_t\rangle=|\langle 0|\psi_t\rangle|^2+\sum_{n\neq 0}
|\langle n|\psi_t\rangle|^2\ .
\end{equation}
Consequently, due to eq.\ (2.5) we must have
\begin{equation}
\frac{1}{\tau_0}\simeq\sum_{n\neq 0}\frac{1}{T_n}\ ,
\end{equation}
in agreement with eqs.\ (2.8,10).

\section{Two bosonic excitons}

Let us first follow a procedure, quite often found in the literature,
which assumes that composite excitons can be validly replaced by
elementary bosonic excitons in the low density limit. Their creation
operators are then such that
\begin{equation}
\left[\overline{B}_i,\overline{B}_j^\dag\right]=\delta_{ij}\ ,
\end{equation}
provided that the composite character of the excitons is
included in the Hamiltonian through the matrix elements
of an appropriate phenomenological exciton-exciton
interaction. The semiconductor Hamiltonian is then replaced by an
effective bosonic Hamiltonian
$H_\mathrm{eff}=H_\mathrm{x}+V_\mathrm{xx}$, where the one-body part
reads
\begin{equation}
H_\mathrm{x}=\sum_iE_i\,\overline{B}_i^\dag\,
\overline{B}_i\ ,
\end{equation}
while the exciton-exciton potential is written as
\begin{equation}
V_\mathrm{xx}=\frac{1}{2}\sum_{mnij}V_{mnij}\,\overline{B}_m^\dag\,
\overline{B}_n^\dag\,\overline{B}_i\,\overline{B}_j\ ,
\end{equation}
with $V_{mnij}=V_{nmij}=V_{mnji}$ due to possible changes in the bold indices, while
$V_{mnij}=V_{ijmn}^\ast$ in order to insure hermiticity,
$H_{\mathrm{eff}}=H_{\mathrm{eff}}^\dag$.

As, for Hamiltonian eigenstates, $\delta_{mi}=\langle m|i\rangle=\langle
v|\overline{B}_m\overline{B}_i^\dag|v\rangle$, we have in the two-exciton subspace, due to eq.\
(3.1),
\begin{equation}
\langle v|\overline{B}_m\,\overline{B}_n\,\overline{B}_i^\dag\,
\overline{B}_j^\dag|v\rangle=\langle
v|\overline{B}_m\,(\overline{B}_i^\dag\,\overline{B}_n+\delta_{in})\,
\overline{B}_j^\dag|v\rangle=
\delta_{mi}\,\delta_{nj}+\delta_{mj}\,
\delta_{ni}\ .
\end{equation}
Consequently, the normalized two-exciton
states are given by
\begin{equation}
|\overline{\phi}_{ij}\rangle=\frac{\overline{B}_i^\dag\,
\overline{B}_j^\dag
|v\rangle}{\sqrt{1+\delta_{ij}}}=|\overline{\phi}_{ji}\rangle\ .
\end{equation}

Transition rates towards other
two-exciton states exist because $|\overline{\phi}_{ij}\rangle$ is not
eigenstate of the Hamiltonian, so that this state evolves with time. In
this time evolution, the
$|\overline{\phi}_{ij}\rangle$ state gets non-zero projections over
$|\overline{\phi}_{mn}\rangle$ with $(mn)\neq (ij)$. For simplicity,
let us study the scattering of two excitons in the same state
$0 \,{\bf \equiv}\, (\nu_0,\v Q_0)$, towards any other two-exciton state.
The normalized initial state is then
$|\overline{\psi}_{t=0}\rangle=|\overline{\phi}_{00}\rangle=(1/\sqrt{2})\overline{B}_0^{\dag
  2}|v\rangle$. By noting that, due to eqs.\ (3.2, 3.3),
\begin{equation}
H_\mathrm{eff}\,\overline{B}_p^\dag\overline{B}_q^\dag|v\rangle=
(E_p+E_q)\,\overline{B}_p^\dag\overline{B}_q^\dag|v\rangle+\sum_{mn}V_{mnpq}\,\overline{B}_m^\dag\,
\overline{B}_n^\dag|v\rangle\ ,
\end{equation}
we get, from eq.\ (3.4), the expectation value of the Hamiltonian in
this $|\overline{\psi}_0\rangle$ initial state as
\begin{equation}
\langle
H_\mathrm{eff}\rangle_{0}=\langle\overline{\phi}_{00}|H_\mathrm{eff}|
\overline{\phi}_{00}\rangle
=2\,E_0+V_{0000}
\end{equation}

Using eqs.\ (2.2, 3.6, 3.7), we then find
\begin{equation}
P_\perp\,H_\mathrm{eff}\,\overline{B}_0^{\dag
2}|v\rangle=(H_\mathrm{eff}-\langle H_\mathrm{eff}\rangle_0)
\overline{B}_0^{\dag 2}|v\rangle=(V_\mathrm{xx}-V_{0000})
\overline{B}_0^{\dag 2}|v\rangle=\sum_{mn\neq 00}V_{mn00}
\overline{B}_m^\dag\overline{B}_n^\dag|v\rangle\ ,
\end{equation}
since, due to eqs.\ (3.3,3.4) we do have
\begin{equation}
V_{\mathrm{xx}}\,\overline{B}_0^{\dag
2}|v\rangle=\frac{1}{2}\sum_{mnij}V_{mnij}
\overline{B}_m^\dag \overline{B}_n^\dag|v\rangle\langle v|\overline{B}_i
\overline{B}_j\overline{B}_0^{\dag 2}|v\rangle=\sum_{mn}V_{mn00}
\overline{B}_m^\dag\overline{B}_n^\dag|v\rangle\ .
\end{equation}
To first order in the interactions, the state change
$|\tilde{\overline{\psi}}_t\rangle$,  given by eq.\ (2.3), is obtained by
replacing $\hat{H}$ by its zero order contribution, namely
$H_\mathrm{x}-2E_0$. This leads to
\begin{equation}
|\tilde{\overline{\psi}}_t\rangle=\sum_{mn\neq 00}F_t(E_m+E_n-2E_0)V_{mn00}
\frac{\overline{B}_m^\dag\overline{B}_n^\dag|v\rangle}{\sqrt{2}}
+O(V^2)\ .
\end{equation}
It is clear from the start that, for physically relevant --- not too large --- momentum ${\bf
Q}_0$, energy conservation imposes the scattered states $(m,n)$ to belong to
the same relative motion subspace $\nu_0$ as the initial excitons.
We then note that momentum conservation in the Coulomb interactions
leading to the exciton-exciton scattering
$V_{mn00}$ imposes the
$(m,n)$ states forming $|\tilde{\overline{\psi}}_t\rangle$ to be such that
$m=(\nu_0,\v Q_0+\v q)$ and $n=(\nu_0,\v Q_0-\v q)$, the momentum
transfer $\v q$ differing from $\v 0$, in order for $(m,n)$ to differ from
$(0,0)$. Consequently, the scattered states are made of different
excitons,
$m\neq n$.

The transition rate towards a (normalized) two-bosonic-exciton state
$|\overline{\phi}_{ij}
\rangle$ with $(i,j)\neq (0,0)$ is thus given by
\begin{equation}
\frac{t}{\overline{T}_{ij}}\simeq|\langle\overline{\phi}_{ij}|\overline{\psi}_t\rangle|^2=
|\langle\overline{\phi}_{ij}|
\tilde{\overline{\psi}}_t\rangle|^2\ ,
\end{equation}
as $\langle\overline{\phi}_{ij}|\overline{\phi}_{00}\rangle=0$ for
$(i,j)\neq (0,0)$, due to eq.\ (3.4). This leads to
\begin{equation}
\frac{t}{\overline{T}_{ij\neq 00}}\simeq\frac{1}{2(1+\delta_{ij})}\left|
\sum_{mn\neq 00}F_t(E_m+E_n-2E_0)\,V_{mn00}\,\langle v|\overline{B}_i
\overline{B}_j\overline{B}_m^\dag\overline{B}_n^\dag|v\rangle\right|^2\
.
\end{equation}
We then note that, due to energy and
momentum conservation, this transition rate differs from zero for
$i\neq j$ only. So that eqs.\ (3.4,2.9) give the transition rate
from $|\overline{\phi}_{00}\rangle$ to $|\overline{\phi}_{ij\neq 00}
\rangle$ as
\begin{eqnarray}
\frac{1}{\overline{T}_{ij\neq 00}}&\simeq & 4\,\pi\
|V_{ij00}|^2\,\delta_t (E_i+E_j-2E_0)\nonumber \\
&=&2\pi\left|\langle\overline{\phi}_{ij}
|V_{\mathrm{xx}}|\overline{\phi}_{00}\rangle\right|^2\delta_t(E_i+E_j-2E_0)\
,
\end{eqnarray}
in agreement with the well known expression of the Fermi golden rule.

If we now turn to the bosonic-exciton lifetime, we see from eq.\ (3.10)
that
$\langle\overline{\phi}_{00}|\tilde{\overline{\psi}}_t\rangle=0(V^2)$, so that, to
second order in $V$, we have \newpage
\begin{eqnarray}
\frac{t}{\overline{\tau_0}}&\simeq &
\langle\tilde{\overline{\psi}}_t|\tilde{\overline{\psi}}_t\rangle
\nonumber \\ &\simeq & \frac{1}{2}
\sum_{mn\neq 00,ij\neq 00}F_t^\ast(E_i+E_j-2E_0)F_t(E_m+E_n-2E_0)
V_{ij00}^\ast V_{mn00}\,\langle v|\overline{B}_i\overline{B}_j
\overline{B}_m^\dag\overline{B}_n^\dag
|v\rangle\nonumber
\\ &\simeq & t\ 2\,\pi\,\sum_{ij\neq 00}|V_{ij00}|^2\,
\delta_t(E_i+E_j-2E_0)\ .
\end{eqnarray}
We thus end with
\begin{equation}
\frac{1}{\overline{\tau}_0}\simeq\frac{1}{2}\sum_{ij\neq00}\frac{1}{
\overline{T}_{ij}}=
\sum_{\mathrm{couples}(i,j)\neq(0,0)}\frac{1}{\overline{T}_{ij}}\ .
\end{equation}

We can also recover the above link between lifetime and transition rates
by using the closure relation for two-bosonic exciton
states. Since any 2-boson state $|\overline{\Phi}\rangle$ can be rewritten
as
\begin{equation}
|\overline{\Phi}\rangle=\frac{1}{2!}\sum_{ij}\overline{B}_i^\dag\,
\overline{B}_j^\dag|v\rangle\langle v|\overline{B}_i\overline{B}_j|
\overline{\Phi}\rangle\ ,
\end{equation}
--- easy to check from eq.\ (3.4), --- the closure relation for
elementary bosons reads in terms of the \emph{normalized} bosonic exciton
states $|\overline{\phi}_{ij}\rangle$ as
\begin{equation}
I=\frac{1}{2}\sum_{ij}\overline{B}_i^\dag\,\overline{B}_j^\dag|v\rangle
\langle
v|\overline{B}_i\overline{B}_j=\frac{1}{2}\sum_{ij}(1+\delta_{ij})
|\overline{\phi}_{ij}\rangle\langle\overline{\phi}_{ij}|
=\sum_i|\overline{\phi}_{ii}\rangle
\langle\overline{\phi}_{ii}|+\frac{1}{2}\sum_{i\neq j}
|\overline{\phi}_{ij}\rangle\langle\overline{\phi}_{ij}|\ .
\end{equation}
If we use this closure relation into $\langle\overline{\psi}_t|\overline{\psi}_t\rangle$, we get
\begin{equation}
1=\langle\overline{\psi}_t|\overline{\psi}_t\rangle=\langle\overline{\psi}_t|\left[\sum_i
|\overline{\phi}_{ii}
\rangle\langle\overline{\phi}_{ii}|+\frac{1}{2}\sum_{i\neq j}
|\overline{\phi}_{ij}\rangle\langle\overline{\phi}_{ij}|\right]
|\overline{\psi}_t\rangle\ .
\end{equation}
We now remember that
$|\overline{\psi}_t\rangle=|\overline{\phi}_{00}\rangle+|\tilde{\overline{\psi}}_t\rangle$, where 
$|\tilde{\overline{\psi}}_t\rangle$ contains terms in $mn\neq 00$ with $m\neq n$ due to momentum
conservation in the Coulomb scatterings producing the time evolution. This leads to
$\langle\overline{\phi}_{ii}|\overline{\psi}_t\rangle=0$ for $i\neq 0$, while the sum over $i\neq
j$ can be replaced by a sum over $(ij\neq 00)$. So that we end with the lifetime of the
$|\overline{\phi}_{00}\rangle$ state given by
\begin{equation}
\frac{t}{\overline{\tau}_0}\simeq
1-|\langle\overline{\phi}_{00}|\overline{\psi}_t\rangle|^2 =\frac{1}{2}\sum_{ij\neq
00}|\langle\overline{\phi}_{ij}|\overline{\psi}_t
\rangle|^2\simeq\frac{t}{2}\sum_{ij\neq 00}\frac{1}{\overline{T}_{ij}}\ ,
\end{equation}
in agreement with eq.\ (3.15).

\section{Two composite excitons}

We now turn to real physics with excitons made of
electron-hole pairs. The semiconductor Hamiltonian,
$H=H_\mathrm{e}+H_\mathrm{h}+\mathcal{V}_\mathrm{ee}+\mathcal{V}
_\mathrm{hh}+\mathcal{V}_\mathrm{eh}$,
reads in terms of electron and hole creation
operators
$a_{\v k}^\dag$ and $b_{\v k}^\dag$, with any two-pair state \emph{a
priori} written as a sum of products of two free pair operators
$a_{\v k_e}^\dag b_{\v k_h}^\dag$. As in the low density limit,
these pairs form excitons, the
representation of two-pair states in terms of two excitons is clearly
the relevant one in this limit, the way to go from the free-pair representation to the exciton
representation being given by eq.\ (A5), in appendix A.

Although physically relevant, this exciton representation raises
some major difficulties which have
to be faced and handled if we want to use it safely. The first one comes
from the fact that the
$N$-exciton states are not orthogonal for $N=2$ already. Indeed, by using
eqs.\ (1.2, 1.3) and by noting that $D_{ni}|v\rangle=0$, which readily
follows from eq.\ (A.3), we find
\begin{eqnarray}
\langle v|B_mB_nB_i^\dag
B_j^\dag|v\rangle &=& \langle v|B_m(B_i^\dag B_n+
\delta_{ni}-D_{ni})B_j^\dag|v\rangle\nonumber
\\ &=& \delta_{mi}\,\delta_{nj}
+\delta_{mj}\,\delta_{ni}-\lambda\left(^{n\ j}_{m\ i}\right)
-\lambda\left(^{m\ j}_{n\ i}\right)\ ,
\end{eqnarray}
which differs from the scalar product of two-boson-exciton states given in eq.\ (3.4) by the
presence of the two $\lambda$ terms. Because of these exchange parameters
which originate from the composite character of the excitons via
the deviation-from-boson operator $D_{mi}$, we see that, unlike boson excitons, the two-exciton
states are never orthogonal, even if the excitons $(m,n)$
are different from $(i,j)$.

Another difficulty, which is related to the above one, comes from the
fact that there is an infinite number of representations of
a given state $|\psi\rangle$ in terms of excitons. This difficulty is directly linked
to the identity (1.6). Indeed, if we know one representation of a state
$|\psi\rangle$ in terms of excitons, we can produce an infinite
number of equally valid representations by using this equation (1.6). Indeed
\begin{eqnarray}
|\psi\rangle &=&\sum_{ij}\psi_{ij}\,B_i^\dag
B_j^\dag|v\rangle\nonumber
\\ &=& \sum_{mn}\left(-\sum_{ij}\lambda\left(^{n\ j}_{m\
i}\right)\,\psi_{ij}\right) B_m^\dag B_n^\dag|v\rangle\nonumber
\\ &=&
\sum_{ij} \left(x\,\psi_{ij}-y\,\sum_{mn}\lambda\left(^{j\ n}_{i\
m}\right)\,
\psi_{mn} \right)\,B_i^\dag B_j^\dag|v\rangle\nonumber\\
&=& \sum_{ij}\psi_{ij}(x,y)\,B_i^\dag B_J^\dag|v\rangle\ ,
\end{eqnarray}
where $x$ and $y$ are two arbitrary constants such that $x+y=1$. Starting from one representation
of $|\psi\rangle$ in terms of excitons, we can thus construct an infinite number of prefactors
$\psi_{ij}(x,y)$ for the same state $|\psi\rangle$.

In spite of these difficulties, the exciton
representation is the relevant one at low density, the system
being closer to excitons than to free pairs. Consequently, it is necessary to learn
how to work with these exciton states properly and to cope with all the
underlying problems associated to them.

In this work, we are interested in the exciton-exciton transition rate.
Such a transition rate exists because the two-exciton states are not
eigenstates of the semiconductor Hamiltonian, so that they evolve with
time. Due to eq.\ (4.1), these normalized two-exciton states read,
instead of eq.\ (3.5),
\begin{equation}
|\phi_{ij}\rangle=\frac{B_i^\dag B_j^\dag|v\rangle}{\sqrt{1+\delta_{ij}
-\lambda\left(^{j\ j}_{i\
i}\right)-\lambda\left(^{i\ j}_{j\
i}\right)}}\ .
\end{equation}
As for bosonic excitons, let us take as initial state two excitons in the
same state
$0 \,{\bf \equiv}\, (\nu_0,\v Q_0)$,
\begin{equation}
|\psi_0\rangle=|\phi_{00}\rangle=\frac{B_0^{\dag 2}|v\rangle}
{\sqrt{2-2\lambda\left(^{0\ 0}_{0\
0}\right)}}\ .
\end{equation}
In order to calculate its time change $|\tilde{\psi}_t\rangle$ given
in eq.\ (2.3), we first need to calculate $H\,B_0^{\dag 2}|v\rangle$.
Using eqs.\ (1.7, 1.8) and noting that
$V_i^\dag|v\rangle=0$, which follows from eq.\ (1.7), we readily find
\begin{equation}
(H-2E_0)B_0^{\dag 2}|v\rangle=\sum_{mn}
\xi^\mathrm{dir}\left(^{n\ 0}_{m\
0}\right)\,B_m^\dag B_n^\dag|v\rangle\ .
\end{equation}

Before going further, let us mention an additional difficulty with these
exciton states, which is another aspect of the one leading to eq.\
(4.2). Making use of eq.\ (1.6), one sees that the RHS of
the above equation is unchanged if we
replace $\xi^\mathrm{dir}\left(^{n\ 0}_{m\
0}\right)$ by
$-\xi^\mathrm{in}\left(^{n\ 0}_{m\
0}\right)$, since
$\xi^\mathrm{dir}$ and $\xi^\mathrm{in}$ are related by eq.\ (1.10).
This shows, using the same procedure as in eq.\ (4.2), that we can, in eq.\ (4.5), replace
$\xi^\mathrm{dir}\left(^{n\ 0}_{m\ 0}\right)$ by $\xi\left(^{n\ 0}_{m\ 0}\right)$, with
\begin{equation}
\xi\left(^{n\ j}_{m\
i}\right) \equiv a\,\xi^\mathrm{dir}\left(^{n\ j}_{m\
i}\right)-b\,\xi^\mathrm{in}\left(^{n\ j}_{m\
i}\right)\ ,
\end{equation}
where $a$ and $b$ are two arbitrary constants such that $a+b=1$, with
$a$ and $b$ chosen real for simplicity. Consequently, the scattering of two
excitons into two other excitons may seem somewhat arbitrary. 

We can
note that
\begin{equation}
\xi'\left(^{n\ j}_{m\ i}\right)=-\sum_{pq}\lambda\left(^{n\ q}_{m\ p}\right)\,\xi\left(^{q\ j}_{p\
i}\right)=-a\xi^\mathrm{in}\left(^{n\ j}_{m\ i}\right)+b
\xi^\mathrm{dir}\left(^{n\ j}_{m\ i}\right)\ ,
\end{equation}
which follows from eq.\ (1.10) and eq.\ (C.1) in appendix C. Consequently, the scatterings
$\xi\left(^{n\ j}_{m\ i}\right)$ which stay unchanged by carrier
exchanges, \emph{i.\ e.}, by, in eq.\ (4.5), rewriting $B_m^\dag B_n^\dag$ according to eq.\
(1.6), are such that
$\xi\left(^{n\ j}_{m\ i}\right)=\xi'\left(^{n\ j}_{m\ i}\right)$. They thus correspond to
$a=b=1/2$. This may lead to think that, among all the possible exciton-exciton Coulomb
scatterings $\xi\left(^{n\ j}_{m\ i}\right)$, the ones which are physically relevant, due to
possible carrier exchanges between composite excitons, are precisely those stable under
these exchanges, \emph{i.\ e.}, those with
$a=b=1/2$. 

Nevertheless, in order to fully master the apparent
arbitrariness in the exciton-exciton Coulomb scatterings, we are, in the
following, going to perform all calculations with $\xi$'s having
arbitrary $a$ and $b$. This will allow us to see, in details, where this
apparent arbitrariness actually disappears from physical quantities. We will also see that the
scatterings with $a=b=1/2$, even if they may appear as physically nicer, are not any better than
the other ones, with respect to the correctness of the lifetime and transition rates.

Using eq.\ (4.5), we find that the
expectation value of the Hamiltonian in the initial state
$|\phi_{00}\rangle$ is given by
\begin{equation}
\langle H\rangle_0=\langle\phi_{00}|H|\phi_{00}\rangle=2\,E_0+\sum_{mn}
\xi\left(^{n\ 0}_{m\
0}\right)\,\frac{\langle v|B_0^2B_m^\dag B_n^\dag|v\rangle}
{2-2\lambda\left(^{0\ 0}_{0\
0}\right)}\ .
\end{equation}
Using eqs.\ (4.1,4.6,4.7), we first see that the sum in the above
equation does not depend on the arbitrariness of $\xi$. Indeed, we do have
\begin{eqnarray}
\sum_{mn}\langle v|B_pB_qB_m^\dag B_n^\dag|v\rangle\,\xi
\left(^{n\ j}_{m\ i}\right)&=&\left[\xi\left(^{q\ j}_{p\ i}\right)-
\sum_{mn}\lambda\left(^{q\ n}_{p\ m}\right)\,\xi
\left(^{n\ j}_{m\ i}\right)\right]+(p\leftrightarrow q)\nonumber\\
&=&\left[\xi\left(^{q\ j}_{p\ i}\right)+\xi'
\left(^{q\ j}_{p\ i}\right)\right]+(p\leftrightarrow q)\nonumber\\
&=&\left[\xi^\mathrm{dir}\left(^{q\ j}_{p\ i}\right)-\xi^\mathrm{in}
\left(^{q\ j}_{p\ i}\right)\right]+(p\leftrightarrow q)\ ,
\end{eqnarray}
whatever $a$ and $b$ are. Consequently, the Hamiltonian expectation value
does not depend on the particular scattering $\xi$ used to
calculate it, as physically reasonable : 
\begin{equation}
\langle H\rangle_0=2E_0+\frac{\xi^\mathrm{dir}\left(^{0\ 0}_{0\ 0}
\right)-\xi^\mathrm{in}\left(^{0\ 0}_{0\ 0}\right)}{1-\lambda
\left(^{0\ 0}_{0\ 0}\right)}\ .
\end{equation}

In the case of excitons, where the Coulomb repulsion
between identical fermions is equal to the attraction
of different fermions, the direct scattering is such that
$\xi^\mathrm{dir}\left(^{0\ 0}_{0\ 0}\right)=0$, due to eq.\ (B.18).
However, in order for this work to be easily transposed to other cobosons
like the cold gases of atomic physics for which the
interactions between fermions do not have the above symmetry, we will
keep writing this
$\xi^\mathrm{dir}\left(^{0\ 0}_{0\ 0}\right)$ in the following
equations.

Using the value of $\langle H\rangle_0$ given in eq.\ (4.10), it is easy
to see that
$P_{\perp}HB_0^{\dag 2}|v\rangle$ entering the state change
$|\tilde{\psi}_t\rangle$ reads
\begin{equation}
P_\perp\,H\,B_0^{\dag 2}|v\rangle = (H-\langle H\rangle_0)B_0^{\dag 2}
|v\rangle
=\sum_{mn}\xi\left(^{n\ 0}_{m\
0}\right)\,B_m^\dag B_n^\dag|v\rangle-V_{00}
\,B_0^{\dag 2}|v\rangle\ ,
\end{equation}
with $V_{00}$ defined as $V_{00}=\left[\xi^\mathrm{dir}\left(^{0\ 0}_{0\ 0}\right)-
\xi^\mathrm{in}\left(^{0\ 0}_{0\ 0}\right)\right]/\left[1-\lambda
\left(^{0\ 0}_{0\ 0}\right)\right]$. Equation (4.11) makes $P_{\perp}HB_0^{\dag
2}|v\rangle$ linear in Coulomb interaction, as physically expected.

If we now consider the transition rate from the initial state
$|\phi_{00}\rangle$ to another exciton state $|\phi_{ij\neq 00}\rangle$,
a na\"{\i}ve way to define this transition rate would be
\begin{equation}
\frac{t}{T^\mathrm{na\ddot{\i}ve}_{ij}}\simeq|\langle\phi_{ij}|\psi_t\rangle|^2=|\langle\phi_{ij}|
\phi_{00}\rangle+\langle\phi_{ij}|\tilde{\psi}_t\rangle|^2\ .
\end{equation}
We then note that, while in usual problems dealing with transition
rates, the initial and final states are orthogonal, being
eigenstates of the unperturbed Hamiltonian $H_0$, this is not true
for transitions between exciton states. Indeed, due to eq.\ (4.1),
the initial and final states are such that
\begin{equation}
\langle\phi_{ij\neq 00}|\phi_{00}\rangle=-\frac{2\,\lambda\left(
^{j\ 0}_{i\ 0}\right)}
{\sqrt{\left[2-2\lambda\left(
^{0\ 0}_{0\
0}\right)\right]\left[1+\delta_{ij}-\lambda\left(
^{j\ j}_{i\ i}\right)-\lambda\left(
^{i\ j}_{j\ i}\right)\right]}}\neq 0\ .
\end{equation}
However, since this scalar product does not depend on time, while
$|\tilde{\psi}_{t=0}\rangle$ is equal to zero, as seen from eq.\ (2.3),
the RHS of eq.\ (4.12) does not cancel for $t=0$, so that it cannot be
equal to
$t/T_{ij}$. Actually, this nonzero
contribution to the
transition rate is completely unphysical because it corresponds to the
fraction of the $|\phi_{ij\neq 00}\rangle$ state already present in the
initial state, due to the nonorthogonality of composite exciton states.

When we speak of transition rate, we have in mind a transition induced
by the time evolution of the system,
the state change associated to this time evolution being nothing but
$|\tilde{\psi}_t\rangle$. Consequently, the physically relevant expression
of the transition rate from $|\phi_{00}\rangle$ to $|\phi_{ij\neq
00}\rangle$ cannot be eq.\ (4.12), but instead
\begin{equation}
\frac{t}{T_{ij}}=|\langle\phi_{ij}|\tilde{\psi}_t\rangle|^2=\left|
\langle\phi_{ij}|F_t(\hat{H})\,P_{\perp}H|\phi_{00}\rangle\right|^2\ ,
\end{equation}
in which we have used eq.\ (2.3) for $|\tilde{\psi}_t\rangle$. Since, due to eq.\ (4.11),
$P_{\perp}H|\phi_{00}\rangle$ is linear in Coulomb interaction, we just have to take $F_t(\hat{H})$
at zero order in Coulomb interaction, in order to get the transition rate at second
order. However, unlike for bosonic excitons for which a zero order
Hamiltonian
$H_\mathrm{x}$ exists, the zero order in Coulomb interaction is less
clear for composite excitons. Of course, we guess that ``zero order''
means to replace in eq.\ (4.14) $\hat{H}$ by $(E_i+E_j-2E_0)$.
Actually, this can be shown in a clean way. By using the following
identities, rederived in appendix E,
\begin{equation}
\frac{1}{a-H}\,B_i^\dag=\left(B_i^\dag+\frac{1}{a-H}\,V_i^\dag\right)
\,\frac{1}{a-H-E_i}\ ,
\end{equation}
\begin{equation}
e^{-iHt}\,B_i^\dag=B_i^\dag\,e^{-i(H+E_i)t}+W_i^\dag (t)\ ,
\end{equation}
with $W_i^\dag (t)$ given by 
\begin{equation}
W_i^\dag (t)=-\int_{-\infty}^{+\infty}\frac{dx}{2i\pi}\,\frac{e^{-ixt}}
{x-H+i\eta}\,V_i^\dag\,\frac{1}{x-H-E_i+i\eta}\ ,
\end{equation}
it is straightforward to show that
\begin{equation}
F_t(H-a)\,B_i^\dag=B_i^\dag\,F_t(H+E_i-a)+\mathcal{V}_i^\dag (t)\ ,
\end{equation}
where $\mathcal{V}_i^\dag (t)$, equal to
\begin{equation}
\mathcal{V}_i^\dag (t)=\left[F_t(H-a)\,V_i^\dag-e^{iat}\,W_i^\dag (t)
\right]\,\frac{1}{a-H-E_i}\ ,
\end{equation}
gives zero when acting on vacuum, while, due to eq.\ (1.8), it gives a
state which is at least linear in $\xi$ when acting on excitons.
Consequently, since $(H-E_j)\,B_j^\dag|v\rangle=0$, we do end with the
result we guessed, namely
\begin{eqnarray}
F_t(H-\langle H\rangle_0)\,B_i^\dag
B_j^\dag|v\rangle&=&F_t(E_i+E_j-\langle H\rangle_0)\,B_i^\dag
B_j^\dag|v\rangle +\mathcal{V}_i^\dag (t)\,B_j^\dag|v\rangle\nonumber\\
&=&F_t(E_i+E_j-2E_0)\,B_i^\dag B_j^\dag|v\rangle+O(\xi)\ .
\end{eqnarray}

The contribution to the transition rate, quadratic in Coulomb
scatterings, is thus given by
\begin{equation}
\frac{t}{T_{ij\neq 00}}=\frac{\left|F_t(E_i+E_j-2E_0)\,\langle v|B_iB_j
\left\{\sum_{mn}\xi
\left(^{n\ 0}_{m\ 0}\right)\,B_m^\dag B_n^\dag
|v\rangle-V_{00}\,B_0^{\dag 2 }|v\rangle\right\}\right|^2}
{\left[1+\delta_{ij}-\lambda\left(^{j\ j}_{i\ i}\right)-\lambda
\left(^{i\ j}_{j\ i}\right)\right]\left[2-2\lambda\left(^{0\ 0}_{0\ 0}
\right)\right]}\ .
\end{equation}
Due to eq.\ (4.9), the scalar product with the sum in the above equation gives
$2\left[\xi^\mathrm{dir}\left(^{j\ 0}_{i\ 0}\right)-
\xi^\mathrm{in}\left(^{j\ 0}_{i\ 0}\right)\right]$ whatever $a$ and $b$
are, while the scalar product of the other term
reduces to $\left[-2\lambda\left(^{j\ 0}_{i\ 0}\right)\right]V_{00}$, for
$ij\neq 00$. We then note, either from dimensional arguments or from
precise calculations using eq.\ (1.4), that the exchange parameters
between bound state excitons are of the order of $(a_X/L)^D$. So that
this other term, which is a factor $(a_X/L)^D$ smaller than the sum, can be
neglected in eq.\ (4.21). Since energy and momentum conservation in the
transition rate imposes
$i\neq j$, we end with a transition rate from
$|\phi_{00}\rangle$ to
$|\phi_{ij\neq 00}\rangle$ given by
\begin{equation}
\frac{1}{T_{ij\neq 00}}=
4\,\pi \left|\xi^\mathrm{dir}\left(
^{j\ 0}_{i\ 0}\right)-
\xi^\mathrm{in}\left(
^{j\ 0}_{i\ 0}\right)\right|^2\,\delta_t(E_i+E_j-2E_0)\ .
\end{equation}
Note that the above expression is formally identical to the one for
bosonic excitons obtained in eq.\ (3.13). We will come back to this
similarity later on.

Let us now turn to the lifetime of the state $|\phi_{00}\rangle$. We can
obtain it from eq.\ (2.6). By replacing $F_t(\hat{H})$ by
its zero order contribution, we first note, using eq.\ (2.3) with $F_t$
acting on the left, that
\begin{equation}
\langle\phi_{00}|\tilde{\psi}_t\rangle\simeq
F_t(0)\langle\phi_{00}|P_{\perp} H\phi_{00}\rangle
\end{equation}
is second order in $\xi$ since $\langle\phi_{00}|P_{\perp}=0$.
Consequently,
$|\langle\phi_{00}|\tilde{\psi}_t\rangle|^2$ gives a contribution to
the lifetime in $\xi^4$. If we now turn to $\langle\tilde{\psi}_t|
\tilde{\psi}_t\rangle$, it reads, using eq.\ (4.11),
\begin{eqnarray}
\langle\tilde{\psi}_t|\tilde{\psi}_t\rangle=\frac{1}{2-2\lambda
\left(^{0\ 0}_{0\ 0}\right)}\,\langle v|B_0^2HP_{\perp}\left\{
\sum_{mn}|F_t(E_m+E_n-2E_0)|^2\xi\left(^{n\ 0}_{m\ 0}\right)B_m^\dag
B_n^\dag|v\rangle\right.\nonumber\\
\left.-V_{00}F_t(0)B_0^{\dag 2}|v\rangle\right\}+O(\xi^3)\ .
\end{eqnarray}
Since $P_{\perp}B_0^{\dag 2}|v\rangle=0$, we can drop the $B_0^{\dag 2}
|v\rangle$ terms of the bracket, so that
\begin{eqnarray}
\langle\tilde{\psi}_t|\tilde{\psi}_t\rangle &=& \frac{1}{2-2\lambda
\left(^{0\ 0}_{0\ 0}\right)}\,\langle v|B_0^2HP_{\perp}\sum_{mn\neq 00}
B_m^\dag B_n^\dag|v\rangle\left|F_t(E_m+E_n-2E_0)\right|^2\xi\left(
^{n\ 0}_{m\ 0}\right)\nonumber\\
&=& \frac{2\pi\,t}{\left[2-2\lambda\left(^{0\ 0}_{0\ 0}\right)\right]}
\sum_{mn\neq 00}\xi\left(^{n\ 0}_{m\
0}\right)\,\delta_t(E_m+E_n-2E_0)\nonumber\\
&\times& \left[\sum_{ij}\xi\left(^{j\ 0}_{i\ 0}\right)^\ast
\langle v|B_iB_jB_m^\dag B_n^\dag|v\rangle
-V_{00}^\ast\langle v|B_0^2B_m^\dag B_n^\dag|v\rangle\right]\ .
\end{eqnarray}
From eq.\ (4.9), the sum in the bracket gives $[2\xi^\mathrm{dir}
\left(^{n\ 0}_{m\ 0}\right)-2\xi^\mathrm{in}
\left(^{n\ 0}_{m\ 0}\right)]^\ast$, while the second term reduces to
$[-2V_{00}^\ast\lambda\left(^{0\ n}_{0\ m}\right)]$, as $mn\neq
00$, which makes it negligible compared to the first term since
$\lambda\left(^{0\ n}_{0\ m}\right)=O\left(a_X^D/L^D\right)$. By
dropping $\lambda\left(^{0\ 0}_{0\ 0}\right)$ in the denominator of
eq.\ (4.25), for the same reason, we end with
\begin{equation}
\frac{1}{\tau_0}=2\pi\,\sum_{mn\neq 00}\left[\xi^\mathrm{dir}
\left(^{n\ 0}_{m\ 0}\right)-\xi^\mathrm{in}
\left(^{n\ 0}_{m\ 0}\right)\right]^\ast\xi\left(^{n\ 0}_{m\ 0}\right)\,
\delta_t(E_m+E_n-2E_0)+O(\xi^3)\ .
\end{equation}

 From the above expression, we may think that the lifetime depends on the
choice of the arbitrary constants $(a,b)$ entering
$\xi\left(^{n\ 0}_{m\ 0}\right)$. Of course, this is not true ! A simple way
to see it is to come back to the expression (4.11) for $P_{\perp}H
B_0^{\dag 2}|v\rangle$. In it, we can replace $\xi\left(^{n\ 0}_{m\
0}\right)$ by $\xi'\left(^{n\ 0}_{m\ 0}\right)$ defined in eq.\ (4.7),
due to eq.\ (1.6). This leads to replace $\xi\left(^{n\
0}_{m\ 0}\right)$ by $\xi'\left(^{n\ 0}_{m\ 0}\right)$ in
eq.\ (4.26). Since $\xi\left(^{n\ 0}_{m\
0}\right)+\xi'\left(^{n\ 0}_{m\ 0}\right)=\xi^\mathrm{dir}\left(^{n\
0}_{m\ 0}\right)-\xi^\mathrm{in}\left(^{n\ 0}_{m\ 0}\right)$, whatever
$a$ and $b$ are, we find, by taking half of the sum of the two expressions of
$1/\tau_0$, that the
$|\phi_{00}\rangle$ lifetime at second order in the interactions is given by
\begin{equation}
\frac{1}{\tau_0}=\pi\,\sum_{mn\neq
00}\left|\xi^\mathrm{dir}\left(^{n\ 0}_{m\ 0}\right)-
\xi^\mathrm{in}\left(^{n\ 0}_{m\
0}\right)\right|^2\,\delta_t(E_m+E_n-2E_0)\ ,
\end{equation}
whatever the arbitrariness of the scattering $\xi$ used to calculate this lifetime is.

By comparing eqs.\ (4.22) and (4.27), we find that the inverse lifetime
and the scattering rates are related by
\begin{equation}
\frac{1}{\tau_0}=\frac{1}{4}\sum_{ij\neq 00}\frac{1}{T_{ij}}\ .
\end{equation}

Let us end this section by showing how this
link between $\tau_0$ and the $T_{ij}$'s, could have been obtained, along
the idea we have developed in ref.\ [20]. As for the link between
$\overline{\tau}_0$ and the
$\overline{T}_{ij}$'s for bosonic excitons, this derivation relies on a
closure relation which exists for composite exciton states. The existence of such a
closure relation was surprising to us at first, because the two-exciton
states are not eigenstates of any Hamiltonian; they are not even
orthogonal. It is however easy to check that
\begin{equation}
|\Phi\rangle=\frac{1}{4}\sum_{ij}B_i^\dag B_j^\dag|v\rangle
\langle v|B_iB_j|\Phi\rangle\ .
\end{equation}
Indeed, using eqs.\ (4.1) and (1.6), we get
\begin{eqnarray}
\langle v|B_mB_n\left[\frac{1}{4}\sum_{ij}B_i^\dag B_j^\dag|v\rangle
\langle v|B_iB_j|\Phi\rangle\right] &=&
\frac{1}{2}\langle v|B_mB_n|\Phi\rangle-\frac{1}{2}
\langle v|\sum_{ij}\lambda\left(^{n\ j}_{m\
i}\right)B_iB_j|\Phi\rangle\nonumber
\\ &=& \langle v|B_mB_n|\Phi\rangle\ ,
\end{eqnarray}
whatever $(m,n)$ are. Since eq.\ (4.29) is valid for any state
$|\Phi\rangle$, we are led to write
\begin{equation}
I=\frac{1}{4}\sum_{ij}B_i^\dag B_j^\dag|v\rangle\langle v|B_iB_j\ .
\end{equation}
Let us stress that, in contrast to the standard cases, in this closure
relation the exciton states are
\emph{not} normalized.

The above identity, inserted in $\langle\tilde{\psi}_t|\tilde{\psi}_t
\rangle$, leads to
\begin{eqnarray}
\frac{t}{\tau_0}+|\langle\phi_{00}|\tilde{\psi}_t\rangle|^2
=\langle\tilde{\psi}_t|\tilde{\psi}_t\rangle=
\frac{1}{4}\left[|\langle v|B_0^2|\tilde{\psi}_t\rangle|^2+\sum_{ij
\neq 00}|\langle v|B_iB_j|\tilde{\psi}_t\rangle|^2\right]\nonumber\\
=\frac{1}{4}\left\{\left[2-2\lambda\left(^{0\ 0}_{0\ 0}\right)\right]|
\langle\phi_{00}|\tilde{\psi}_t\rangle|^2+\sum_{ij\neq
00}\left[1+\delta_{ij}-\lambda\left(^{j\ j}_{i\ i}\right)-
\lambda\left(^{i\ j}_{j\ i}\right)\right]|\langle\phi_{ij}|\tilde{\psi}
_t\rangle|^2\right\}\ .
\end{eqnarray}
Since the state $|\tilde{\psi}_t\rangle$ is orthogonal to the
initial state $|\phi_{00}\rangle$ at first order in the interactions
(see eq.\ (4.23)), the $\xi^2$ term of the above equation reduces to the
second term, in which $i\neq j$, due to energy and momentum
conservation. Using eq.\ (4.14), we thus end with
\begin{equation}
\frac{1}{\tau_0}=\frac{1}{4}\sum_{ij\neq
00}\frac{1}{T_{ij}}\ ,
\end{equation}
in agreement with eq.\ (4.28).

We thus conclude that the direct calculation of the lifetime and the
one using the closure relation for two-composite-exciton states both
give the same prefactor $1/4$ between $1/\tau_0$ and the
sum of
$1/T_{ij}$'s, instead of 1/2 as for bosonic excitons (see eq.\ (3.15)). As
developed in the next section, this prefactor
$1/4$ has in fact dramatic consequences on a supposedly valid
replacement of composite excitons by bosonic excitons.

\section{Composite boson versus elementary boson: the concept of
``coboson''}

In order to understand the importance of the composite
character of the excitons on the many-body physics of these particles,
it is enlightening to carefully compare the results obtained using
bosonic excitons with similar quantities calculated for composite
excitons. As shown below, \emph{there is no way to produce neither
the correct lifetime nor the correct scattering rates, using an effective
bosonic exciton Hamiltonian, even if we accept this effective Hamiltonian to
be non hermitian}. Consequently, the excitons, as well as any other pairs of fermions,
must be treated for what they really are: composite bosons or ``cobosons'', our work on
exciton-exciton scattering showing the necessity to introduce such an object,
and to construct a specific many-body theory adapted to them.

Even if it is clear that the concept of effective bosonic
Hamiltonian has been mainly introduced for convenience, it is still widely used
because of (i) the very impressive literature which exists on bosonization and which
claims to properly transform the semiconductor Hamiltonian $H$, written with electron and
hole creation operators
$a^\dag$ and $b^\dag$, into an effective Hamiltonian
written in terms of bosonic exciton operators $\overline{B}^\dag$, this
replacement being considered as valid at low density, (ii) the
claimed agreement with experimental results of physical quantities calculated using these
bosonic excitons. The present work shows, as it is well known, that correct
experimental results can be reproduced by incorrect theories.

The procedure used by Haug and Schmitt-Rink [2] to produce a
bosonic exciton effective Hamiltonian is rather famous among semiconductor
physicists, probably because it is the most transparent one. However, the fact
that their procedure is not symmetrical with respect to the ``in'' and
``out'' excitons, should have been a major alert with respect to
its possible correctness. The exciton-exciton scattering Haug and Schmitt-Rink have
produced reads in terms of two of our Coulomb scatterings as
\begin{equation}
V_{mnij}^{(1)}=\xi^\mathrm{dir}\left(^{n\
j}_{m\ i}\right)-\xi^\mathrm{out}\left(^{n\
j}_{m\ i}\right)\ .
\end{equation}
Even if this has not been realized for quite some time, such a scattering has a major
failure : Due to eq.\ (1.11),
\begin{equation}
(V_{mnij}^{(1)})^\ast=\xi^\mathrm{dir}\left(^{j\
n}_{i\ m}\right)-\xi^\mathrm{in}\left(^{j\
n}_{i\ m}\right)
\neq V_{ijmn}^{(1)}\ ,
\end{equation}
so that the associated potential term in the Hamiltonian is not hermitian ! This failure
is easy to miss because physical quantities calculated
using this interaction are expressed in terms of $|V_{mnij}|^2$. However, with
such a failure at the most basic level of the theory, namely in the Hamiltonian itself,
there is
\emph{a priori} no reason to trust the obtained results.

Before going further, let us mention that, if we performed a manipulation
similar to the one done by Haug and Schmitt-Rink, but in a symmetrical
way with respect to the ``in'' and ``out'' excitons, we would end with
\begin{equation}
V_{mnij}^{(2)}=\xi^\mathrm{dir}\left(^{n\
j}_{m\ i}\right)-\frac{\xi^\mathrm{in}\left(^{n\
j}_{m\ i}\right)
+\xi^\mathrm{out}\left(^{n\
j}_{m\ i}\right)}{2}=(V_{ijmn}^{(2)})^\ast\ ,
\end{equation}
which is somewhat better than $V_{mnij}^{(1)}$, as the corresponding Hamiltonian is at
least hermitian.

Actually none of these $V_{mnij}^{(1)}$ or
$V_{mnij}^{(2)}$ nor any other
$V_{mnij}$ can produce the exciton scattering rates and
lifetime correctly, as we now show.

Let us carefully compare the results obtained for bosons and
cobosons, step by step.

(1) The commutator for boson creation operators being just
$\delta_{ij}$, the $\lambda\left(^{n\
j}_{m\ i}\right)$ exchange parameter appearing in
eq.\ (1.2) for cobosons, reduces to zero for elementary bosons.
Consequently, the two last terms of the scalar product $\langle
v|B_mB_nB_i^\dag B_j^\dag| v\rangle$ as given in eq.\ (4.1) disappear
for bosons, in agreement with eq.\ (3.4).

(2) Equation (1.6) has no equivalent for bosons, so that the
decomposition of a state $|\psi\rangle$ is
not unique on coboson states (see eq.\ (4.2)), while it is
unique if we use elementary bosons.

(3) The normalization factor is somewhat more complicated for cobosons
than for bosons (see eqs.\ (3.5) and (4.3)). This is rather
unimportant for $N=2$, since the exchange parameters
$\lambda\left(^{j\ j}_{i\ i}\right)$ between bound states are small
compared to 1. Let us however anticipate in saying that, for large
$N$, this normalization factor has non-trivial consequences in
problems dealing with many-body effects between cobosons. Indeed,
we have shown [22,23] that
\begin{equation}
\langle v|B_0^NB_0^{\dag N}|v\rangle=N!\,F_N\ ,
\end{equation}
where $F_N$ behaves as $F_N\simeq
\exp[-N(\cdots\eta+\cdots\eta^2+\cdots)]$, with
$\eta=N(a_X/L)^D$, while for elementary bosons we just have $\langle
v|\overline{B}_0^N\,\overline{B}_0^{\dag N}|v\rangle=N!$, so that $F_N$
reduces to 1. This change is of importance for large samples since, for a
given density,
\emph{i}.\ \emph{e}., for a given $\eta$, the
product $N\eta$ increases with the sample size, so that it can get
larger than 1, $F_N$ being then exponentially small. In most
semiconductor experiments, $N$ and
$L$ are in fact such that
$N\eta$ is very large, so that in these experiments,
the factor $F_N$ is indeed very different from its bosonic value 1.

(4) By comparing $H_\mathrm{eff}\overline{B}_0^{\dag 2}|v\rangle$ with
$HB_0^{\dag 2}|v\rangle$, as given in eqs.\ (3.6) and (4.5), we see that
the results are identical if we set $V_{mn00}=V_{mn00}^{(3)}$ with
\begin{equation}
V_{mnij}^{(3)}=\xi^\mathrm{dir}\left(^{n\ j}_{m\ i}\right)\ .
\end{equation}
However, due to the carrier exchanges inducing eq.\ (1.6),
it is
\emph{a priori} possible to replace this $\xi^\mathrm{dir}\left(^{n\
j}_{m\ i}\right)$ by
$-\xi^\mathrm{in}\left(^{n\
j}_{m\ i}\right)$. This would lead to take
$V_{mnij}=V_{mnij}^{(4)}$, with
\begin{equation}
V_{mnij}^{(4)}=-\xi^\mathrm{in}\left(^{n\ j}_{m\ i}\right)\ ,
\end{equation}
the physically relevant scattering actually being
\begin{equation}
V_{mnij}^{(5)}=\left[\xi^\mathrm{dir}\left(^{n\ j}_{m\
i}\right)-\xi^\mathrm{in}
\left(^{n\ j}_{m\ i}\right)\right]/2\ ,
\end{equation}
\emph{i.\ e.}, $\xi\left(^{n\ j}_{m\ i}\right)$ for $a=b=1/2$, as it is the only
scattering stable with respect to carrier exchanges (see eq.\ (4.7)).

(5) The situation gets somewhat better for $\langle H\rangle_0$: When
comparing eqs.\ (3.7) and (4.10), we find the same result for bosons
and cobosons if we take $V_{0000}\simeq\xi^\mathrm{dir}\left(^{0\ 0}_{0\
0}\right)-\xi^\mathrm{in}\left(^{0\ 0}_{0\ 0}\right)$, the additional
factor
$\left(1-\lambda\left(^{0\ 0}_{0\ 0}\right)\right)^{-1}$
being negligible for $(a_X/L)\ll 1$. If we
come back to the scatterings
$V_{mnij}^{(1)}$ and $V_{mnij}^{(2)}$ generated by the bosonization
procedure, given in eqs.\ (5.1) and (5.3), we find that they both
lead to the correct $\langle H\rangle_0$ since, for fermions with
Coulomb interaction between them,
$\xi^\mathrm{dir}\left(^{0\ 0}_{0\ 0}\right)=0$, while
$\xi^\mathrm{in}\left(^{0\ 0}_{0\ 0}\right)=\xi^\mathrm{out}\left(^{0\
0}_{0\ 0}\right)$, as shown in appendix D. We can note that the correct
$\langle H\rangle_0$ is also obtained with $V_{mnij}^{(4)}$, but it is
missed with $V_{mnij}^{(3)}$ and $V_{mnij}^{(5)}$, even if this
$V_{mnij}^{(5)}$ can appear at first as the physically relevant
scattering, due to its stability with respect to carrier exchanges.

(6) If we now turn to
$P_\perp H_\mathrm{eff}\overline{B}_0^{\dag 2}|v\rangle$ and $P_\perp
HB_0^{\dag 2}|v\rangle$ as given in eqs.\ (3.8) and (4.11), the results
are identical if we take
$V_{mn00}=\xi\left(^{n\ 0}_{m\ 0}\right)$, with $\xi\left(^{n\ 0}_{m\
0}\right)$ given by eq.\ (4.6). In view of points (4,5,6), a reasonable
choice for
$V_{mnij}$ which fulfils these 3 points appears at this stage, to be
$V_{mnij}=V_{mnij}^{(4)}$,
\emph{i.\ e.,} $\xi\left(^{n\ j}_{m\
i}\right)$ given in eq.\ (4.6), with $a=0$.

(7) If we now compare the scattering rates $1/T_{ij}$ for bosons and
cobosons as given in eqs.\ (3.13) and (4.22), we find that they are equal
if we take for $V_{mnij}$ the quantity
\begin{equation}
V_{mnij}^{(6)}=\xi^\mathrm{dir}\left(^{n\ j}_{m\
i}\right)-\xi^\mathrm{in}\left(^{n\ j}_{m\
i}\right)\ ,
\end{equation}
which is twice the scattering $V_{mnij}^{(5)}$.

Actually, neither $P_\perp H|\psi_0\rangle$ nor
$|\tilde{\psi}_t\rangle$ are physical quantities, so that there are no
strong physical reasons to enforce these quantities to be the same for
bosons and cobosons, by taking $V_{mnij}$ as $V_{mnij}^{(4)}$. In this 
respect, we can note that
$V_{mnij}^{(6)}$ would also produce the correct $\langle H\rangle_0$,
which is rather nice as $\langle H\rangle_0$ is a physical
quantity, being the expectation value of the energy in the initial
state.

This leads us to conclude that the best choice for
$V_{mnij}$, as enforced by the obtention of correct values for the two
physical quantities $1/T_{ij}$ and $\langle H\rangle_0$, appears to be
$V_{mnij}^{(6)}$.

(8) \emph{This best choice is however physically unacceptable} because,
as for
$V_{mnij}^{(1)}$, we have $\left[V_{mnij}^{(6)}\right]^\ast\neq
V_{ijmn}^{(6)}$, so that the corresponding bosonic Hamiltonian would be
non hermitian.

(9) \emph{Another major difference between elementary bosons and cobosons comes from
the link between the inverse lifetime $1/\tau_0$ and the scattering
rates
$1/T_{ij}$}: When comparing eq.\ (3.18) with eq.\ (4.28), we see that
the prefactor 1/2 is transformed into a prefactor 1/4. This factor of 2
change is not unimportant ! It proves, in a transparent way, that it is
impossible to construct a set of $V_{mnij}$'s giving \emph{both} the
lifetime and the scattering rate correctly: This destroys a nice physicist
dream of finding  an effective bosonic
Hamiltonian for excitons, valid for every problem !

This factor of 2 change can be traced back to the closure relations for
bosons and cobosons (see eqs.\ (3.16) and (4.31)). Let us write them
again, as they are crucial in this problem:
$$\overline{I}=\frac{1}{2}\sum_{ij}\overline{B}_i^\dag\,\overline{B}_j^\dag|v
\rangle\langle v|\overline{B}_i\,\overline{B}_j\ ,$$
\begin{equation}
I=\frac{1}{4}\sum_{ij}B_i^\dag\,B_j^\dag|v\rangle\langle v|B_i\,B_j
\ .
\end{equation}
As fully clear from its derivation -- done at the end of section 4 --,
the additional factor 1/2 in the closure relation for cobosons comes
from the fact that, while boson
states are orthogonal, exciton states are
\emph{not}, due to their composite nature. This additional
factor 1/2 is thus an insidious signature of the composite nature of
the cobosons, \emph{i}.\ \emph{e}., of Pauli exclusion between their
components: There is no way to forget this composite nature through an
unique ``dressed exciton-exciton scattering'', valid for everything,
whatever its value is.

\section{Quantitative comparison of the various scatterings}

Let us end this discussion on composite excitons versus boson excitons,
by a \emph{quantitative} comparison of the various scattering rates
appearing in this problem. These scattering rates are based on the four
elementary scatterings which appear in a correct approach to the exciton
many-body physics, namely $\xi^\mathrm{dir}$, $\xi^\mathrm{in}$,
$\xi^\mathrm{out}$ and $\mathcal{E}$. Let us reconsider them quantitatively
in the particular case of scatterings between two ground state excitons in 2D
quantum wells, the two ``in'' excitons having the same center of mass momentum
$\v K$, taken equal to zero for simplicity, since the various
scatterings cannot depend on it, due to translational invariance.

Equation (B.17) given in appendix B gives the direct Coulomb scattering
as
\begin{equation}
\xi^\mathrm{dir}\left(^{\nu_0,-\v Q\ \ \nu_0,\v 0}_{\,\, \nu_0,\v Q\ \ \
\nu_0,\v
0}\right)=\xi^\mathrm{dir}(Q)=V_Q\left|\gamma(\alpha_hQ)-\gamma(-\alpha_eQ)
\right|^2\ ,
\end{equation}
with $\gamma(Q)=\langle\nu_0|e^{i\v Q.\v r}|\nu_0\rangle$. Using the normalized ground
state wave function for 2D excitons given by
\begin{equation}
\langle\v r|\nu_0\rangle=e^{-2r/a_X}\,\sqrt{8/\pi a_X^2}\ ,
\end{equation}
and the 2D Coulomb potential $V_Q=2\pi e^2/L^2Q$, this direct Coulomb
scattering can be calculated analytically. It reads as
\begin{eqnarray}
\xi^\mathrm{dir}(Q)&=&\frac{e^2}{a_X}\,\left(\frac{a_X}{L}\right)^2\,
\tilde{\xi}^\mathrm{dir}(Qa_X)\nonumber\\
\tilde{\xi}^\mathrm{dir}(\tilde{Q})&=&
\frac{2\pi}{\tilde{Q}}\,\left[g\left(\frac{\alpha_h\tilde{Q}}{2}\right)-
g\left(\frac{\alpha_e\tilde{Q}}{2}\right)\right]^2\ ,
\end{eqnarray}
where $g(q)=[1+q^2/4]^{-3/2}$.

We see from eq.\ (6.3) that $\xi^\mathrm{dir}(Q)=0$ for $Q=0$ and
$Q\rightarrow \infty$. We also see that $\xi^\mathrm{dir}(Q)$ stays equal
to zero for $\alpha_e=\alpha_h$, \emph{i.\ e.}, for $m_h=m_e$, its maximum
value being obtained for $\alpha_h=1$, \emph{i.\ e.}, for $m_h\gg m_e$.
Fig.2 shows the behavior of $\tilde{\xi}^\mathrm{dir}(\tilde{Q})$ for
$m_h\gg m_e$ and $m_h=2m_e$.

If we now turn to the Pauli scattering associated with carrier exchanges defined in eq.\
(1.12), we find, from eq.\ (A.7),
\begin{equation}
\mathcal{E}\left(^{\nu_0,-\v Q\ \ \nu_0,\v 0}_{\,\, \nu_0,\v Q\ \ \
\nu_0,\v 0}\right)=\mathcal{E}(Q)=2\,\frac{Q^2}{2M_X}\sum_{\v k}
\langle\nu_0|\v k+\frac{\beta\v Q}{2}\rangle
\langle\nu_0|\v k-\frac{\beta\v Q}{2}\rangle
\langle\nu_0|\v k+\frac{\v Q}{2}\rangle
\langle\nu_0|\v k-\frac{\v Q}{2}\rangle ,
\end{equation}
with $M_X=m_e+m_h$ and $\beta=\alpha_h-\alpha_e$.

Using the normalized ground state wave function in momentum space for 2D excitons,
\begin{equation}
\langle\v k|\nu_0\rangle=\sqrt{2\pi}\,(a_X/L)\,[1+k^2a_X^2/4]^{-3/2}\ ,
\end{equation}
this Pauli scattering can be reduced to a second order integral
\begin{eqnarray}
\mathcal{E}(Q)&=&\frac{e^2}{a_X}\,\left(\frac{a_X}{L}\right)^2\,
\tilde{\mathcal{E}}(Qa_X)\nonumber\\
\tilde{\mathcal{E}}(\tilde{Q})&=&
\alpha_e\
\alpha_h\,\tilde{Q}^2\,\int_0^{+\infty} pdp\,\int_0^{2\pi} d\theta\,
f(p,\theta;\beta\tilde{Q}/2)\,f(p,\theta;\tilde{Q}/2)\nonumber\\
f(p,\theta;K)&=&\left[\left(1+\frac{p^2+K^2}{4}\right)^2-
\left(\frac{pK\cos\theta}{2}\right)^2\right]^{-3/2}\ .
\end{eqnarray}
We see from eq.\ (6.6) that $\mathcal{E}(Q)$ is equal to zero for $Q=0$ and
$Q\rightarrow \infty$. We also see that it stays equal to zero for
$\alpha_e=0$, \emph{i.\ e.}, for $m_h\gg m_e$. The numerical evaluation of
$\tilde{\mathcal{E}}(\tilde{Q})$ for $m_h=m_e$ and $m_h=2m_e$ is
shown in Fig.3.

As rederived in appendix D, this Pauli scattering $\mathcal{E}(Q)$ is
nothing but the difference between the two exchange Coulomb scatterings,
$\xi^\mathrm{in}(Q)-\xi^\mathrm{out}(Q)$, more precisely, the difference
between their electron-hole contributions (see eq.\ (D.1)), since the
electron-electron and hole-hole contributions to $\xi^\mathrm{in}$ and
$\xi^\mathrm{out}$ are identical (see eq.\ (C.4)). Using eqs.\ (C.8,9),
these partial contributions between different fermions read
\begin{eqnarray}
\xi^\mathrm{in}_{\neq}(Q)=\xi^\mathrm{out}_{\neq}(Q)+\mathcal{E}(Q)=
\xi^\mathrm{in}_{eh'}
\left(^{\nu_0,-\v Q\ \ \nu_0,\v 0}_{\,\, \nu_0,\v Q\ \ \
\nu_0,\v 0}\right)+\xi^\mathrm{in}_{he'}
\left(^{\nu_0,-\v Q\ \ \nu_0,\v 0}_{\,\, \nu_0,\v Q\ \ \
\nu_0,\v 0}\right)\nonumber\\
=\sum_{\v k}\left(2\varepsilon_{\nu_0}-\epsilon_{\v k-\frac{\beta\v
Q}{2}}-\epsilon _{\v k+\frac{\beta\v Q}{2}}\right)\,\langle\nu_0|\v
k+\frac{\beta\v Q}{2}\rangle
\langle\nu_0|\v k-\frac{\beta\v Q}{2}\rangle\hspace{0.2cm}\nonumber\\
\times\
\langle\v k+\frac{\v Q}{2}|\nu_0\rangle
\langle\v k-\frac{\v Q}{2}|\nu_0\rangle\ .
\end{eqnarray}
Using the ground state wave function given in eq.\ (6.5), we can rewrite
this scattering in Rydberg units as
\begin{eqnarray}
\xi^\mathrm{in}_{\neq}(Q)&=&-\frac{e^2}{a_X}\,\left(\frac{a_X}{L}\right)^2\,
\tilde{\xi}^\mathrm{in}_{\neq}(Qa_X)\nonumber\\
\tilde{\xi}^\mathrm{in}_{\neq}(\tilde{Q})&=&
\int_0^{+\infty}pdp\int_0^{2\pi}d\theta\,f(p,\theta;\beta\tilde{Q}/2)\,
f(p,\theta;\tilde{Q}/2)
\left[4+p^2+\frac{\beta^2\tilde{Q}^2}{4}
\right]\ .
\end{eqnarray}
The partial ``in'' Coulomb scattering is shown in Fig.4 for $m_h=m_e$,
$m_h=2m_e$  and
$m_h\gg m_e$.

If we now turn to the electron-electron and hole-hole contributions to the
exchange Coulomb scatterings $\xi^\mathrm{in}$ and $\xi^\mathrm{out}$, we find, using eq.\ (C.5),
that the contribution coming from identical fermions is given by 
\begin{eqnarray}
\xi^\mathrm{in}_{=}(Q)=\xi^\mathrm{out}_{=}(Q)=
\xi^\mathrm{in}_{ee}
\left(^{\nu_0,-\v Q\ \ \nu_0,\v 0}_{\,\, \nu_0,\v Q\ \ \
\nu_0,\v 0}\right)+\xi^\mathrm{in}_{hh}
\left(^{\nu_0,-\v Q\ \ \nu_0,\v 0}_{\,\, \nu_0,\v Q\ \ \
\nu_0,\v 0}\right)\hspace{5.5cm}\nonumber\\
=\sum_{\v k}\sum_{\v q\neq\v 0}V_{\v q}\langle\nu_0|\v k+\frac{\beta\v
Q+\v q}{2}\rangle
\langle\nu_0|\v k-\frac{\beta\v Q+\v q}{2}\rangle\hspace{5.6cm}\nonumber\\
\times\left[\langle\v k+\frac{\v Q-\v q}{2}|\nu_0\rangle
\langle\v k-\frac{\v Q-\v q}{2}|\nu_0\rangle+
\langle\v k+\frac{\v Q+\v q}{2}|\nu_0\rangle
\langle\v k-\frac{\v Q+\v q}{2}|\nu_0\rangle\right]\ .
\end{eqnarray}
In Rydberg units, this scattering reads as a fourth order integral
\begin{eqnarray}
\xi^\mathrm{in}_{=}(Q)&=&\xi^\mathrm{out}_{=}(Q)=\frac{e^2}{a_X}\,
\left(\frac{a_X}{L}\right)^2\,\tilde{\mathcal{E}}^\mathrm{in}_{=}(Qa_X)
\nonumber\\
\tilde{\mathcal{E}}^\mathrm{in}_{=}(\tilde{Q})&=&
\int_0^{+\infty}pdp\int_0^{2\pi}d\theta
\int_0^{+\infty}dr\,\int_0^{2\pi}\frac{d\theta'}{2\pi}
\,F(p,\theta;\frac{r}{2},\theta';\frac{\beta\tilde{Q}}{2})\nonumber\\
&\
&\hspace{3cm}\times\left[F(p,\theta;-\frac{r}{2},\theta';\frac{\tilde{Q}}{2})+
F(p,\theta;\frac{r}{2},\theta';\frac{\tilde{Q}}{2})\right]\nonumber
\end{eqnarray}
\begin{eqnarray}
F(p,\theta;q,\theta';K)&=&\left\{\left[1+\frac{p^2+K^2+q^2+
2Kq\cos\theta'}{4}\right]^2\right.\hspace{5cm}\nonumber\\
&\ & \hspace{3cm} \left.-\left[\frac{pK\cos\theta+pq\cos(\theta-\theta')}
{2}\right]^2\right\}^{-3/2}.
\end{eqnarray}
Note that, for $q=0$, the function $F(p,\theta;q,\theta';K)$ reduces to the
function
$f(p,\theta;K)$ entering $\tilde{\mathcal{E}}$ and
$\tilde{\xi}^\mathrm{in}_{\neq}$.

The numerical evaluation of the
electron-electron and hole-hole contribution to the exchange Coulomb
scattering is shown in Fig.5, for
$m_h=m_e$, $m_h=2m_e$ and $m_h\gg m_e$.

From these two partial Coulomb exchange scatterings, we can obtain
$\xi^\mathrm{in}(Q)$ through
\begin{equation}
\xi^\mathrm{in}(Q)=\xi^\mathrm{in}_{=}(Q)+\xi^\mathrm{in}_{\neq}(Q)\ ,
\end{equation}
and $\xi^\mathrm{out}(Q)$ through
\begin{equation}
\xi^\mathrm{out}(Q)=\xi^\mathrm{out}_{=}(Q)+\xi^\mathrm{out}_{\neq}(Q)
=\xi^\mathrm{in}(Q)-\mathcal{E}(Q)\ .
\end{equation}
Note that $\xi^\mathrm{in}(Q)=
\xi^\mathrm{out}(Q)$ for $\mathcal{E}(Q)=0$, \emph{i.\ e.}, for $Q=0$ and
$Q\rightarrow \infty$, as well as for $m_h\gg m_e$ whatever the momentum
transfer $Q$ is.

These two exchange scatterings are shown in Figs.\ 6 and 7. We see that they
both cancel for a finite value of $Q$.

 From these elementary scatterings between two cobosons, we can construct
the two linear combinations of these scatterings appearing in the
transition rates of two ground state excitons with same momentum, towards
two ground state excitons with momentum $+\v Q$ and $-\v Q$,
namely $\xi^\mathrm{dir}(Q)-\xi^\mathrm{in}(Q)$ as obtained through the
many-body theory for composite excitons, and
$\xi^\mathrm{dir}(Q)-\xi^\mathrm{out}(Q)$ as derived by the Haug
and Schmitt-Rink incorrect effective Hamiltonian [2]. The correct
effective scattering $\xi^\mathrm{dir}-\xi^\mathrm{in}$ is plotted in Fig.8 for
$m_h=m_e$, $m_h=2m_e$ and $m_h\gg m_e$, while the two effective scatterings
$\xi^\mathrm{dir}-\xi^\mathrm{in}$ and $\xi^\mathrm{dir}-\xi^\mathrm{out}$
are plotted in Fig.9, to allow an easy comparison of them. We
see that these two effective scatterings have a similar behavior,
as possible to guess from physical arguments, their values
being however significantly different except for $m_h\gg m_e$. We also see that these
effective scatterings both cancel for a finite value of the momentum transfer $Q$, the scattering
rate of the associated process being then infinite. Let us stress that this somewhat unexpected
result is obtained within the Born
approximation --- through the use of the Fermi golden rule. It might be specific to this
approximation and disappear when higher order terms in Coulomb interaction
are taken into account. However, it is also quite possible that this cancellation survives
to all order in Coulomb interaction, since similar situations are known to
occur, for example in atomic physics.

\section{Conclusion}

In the present work (paper I) on the exciton-exciton scattering, we concentrate
on the importance of the exciton composite nature. We show
that there is no way to get rid of this composite nature, by replacing the
excitons by elementary bosons with Coulomb interaction dressed by carrier
exchange, whatever is the way we dress it. For this purpose, we have here studied the
problem of the lifetime and scattering rates of just two excitons, without spin
degree of freedom.

While this paper is written in terms of excitons,
the obtained results can be generalized to other composite
bosons such as the ones found in cold gases much studied these days in 
atomic physics. Since the problems raised by replacing 
composite excitons by elementary excitons are generic, we are led to believe that
fermion exchange between composite atoms should also play a significant role in the physics
of these systems, as well as in the physics of other composite bosons. 

The forthcoming paper II will be devoted to the importance of the spin
degree of freedom in the scatterings of just two excitons and to the
resulting polarization effects, since two bright excitons with opposite
spins scatter into two dark excitons. Finally, paper III will
study the many-body physics associated with these scatterings, through the
time evolution of $N\gg 1$ excitons.

\vspace{1cm}

We wish to thank Marc-Andr\'{e} Dupertuis for a careful study of the manuscript and his valuable
comments.

\newpage

\hbox to \hsize {\hfill \textbf{APPENDICES}
\hfill}

\vspace{0.5cm}

\appendix

\renewcommand{\theequation}{\Alph{section}.\arabic{equation}}

In previous works using our ``commutation technique'' for composite
exciton interactions, we have obtained important results on the various
scatterings between two excitons. The readers interested in this
new many-body theory for cobosons can find useful to have them all
rederived with coherent notations, some of the derivations we here give
being actually simpler than the ones we first proposed.

\section{Exchange parameter}

Using the exciton creation operator $B_i^\dag$ in terms of free pairs given
in eq.\ (1.1), the commutator of two composite exciton operators
appears as
\begin{equation}
\left[B_m,B_i^\dag\right]=\sum_{\v p_e,\v p_h,\v k_e,\v k_h}
\langle\phi_m|\v p_e,\v p_h\rangle\langle\v k_e,\v k_h|\phi_i\rangle
\left[b_{\v p_h}a_{\v p_e},a_{\v k_e}^\dag b_{\v k_h}^\dag\right]\ .
\end{equation}
Since the commutator of two free pair operators is
\begin{equation}
\left[b_{\v p_h}a_{\v p_e},a_{\v k_e}^\dag b_{\v k_h}^\dag\right]
=\delta_{\v p_e,\v k_e}\,\delta_{\v p_h,\v k_h}-\delta_{\v p_h,\v k_h}
\,a_{\v k_e}^\dag a_{\v p_e}-\delta_{\v p_e,\v k_e}\,b_{\v k_h}^\dag
b_{\v p_h}\ ,
\end{equation}
the ``deviation-from-boson operator'' $D_{mi}$ defined in eq.\ (1.3), is
given by
\begin{equation}
D_{mi}=\sum_{\v p_e,\v p_h,\v k_e,\v k_h}\langle\phi_m|\v p_e,\v p_h
\rangle\langle\v k_e,\v k_h|\phi_i\rangle\left(\delta_{\v p_h,\v k_h}
a_{\v k_e}^\dag a_{\v p_e}+\delta_{\v p_e,\v k_e}b_{\v k_h}^\dag
b_{\v p_h}\right)\ .
\end{equation}
Using this expression of $D_{mi}$, we get
\begin{eqnarray}
\left[D_{mi},B_j^\dag\right]=\sum_{\v p_e,\v p_h,\v k_e,\v k_h,\v
k'_e,\v k'_h}\langle\phi_m|\v p_e,\v p_h\rangle\langle\v k_e,\v k_h
|\phi_i\rangle\langle\v k'_e,\v k'_h|\phi_j\rangle\hspace{3cm}\nonumber
\\ \times\ \left(\delta_{\v p_h,\v k_h}\delta_{\v p_e,\v k'_e}a_{\v
k_e}^\dag b_{\v k'_h}^\dag +\delta_{\v p_e,\v k_e}\delta_{\v p_h,
\v k'_h}a_{\v k'_e}^\dag b_{\v k_h}^\dag\right)\ .
\end{eqnarray}
If we now write the free pair operators in terms of exciton operators,
according to
\begin{equation}
a_{\v k_e}^\dag b_{\v k_h}^\dag=\sum_n\langle\phi_n|\v k_e,\v k_h
\rangle\,B_n^\dag\ ,
\end{equation}
easy to check from eq.\ (1.1), we readily get eq.\ (1.2),
in which we have set
\begin{eqnarray}
\lambda \left(^{n\ j}_{m\ i}\right) &=&
\sum_{\v k_e,\v
k_h,\v k'_e,\v k'_h}
\langle\phi_m|\v k_e,\v k'_h\rangle\,\langle\phi_n|\v k'_e,\v k_h
\rangle\,\langle\v k_e,\v k_h|\phi_i\rangle\,
\langle\v k'_e,\v k'_h|\phi_j\rangle\nonumber\\
&=& \lambda \left(^{m\ i}_{n\
j}\right) = \lambda \left(^{j\ n}_{i\ m}\right)^\ast
\ .
\end{eqnarray}
If we turn to real space, this equation gives the expression of
$\lambda\left(^{n\ j}_{m\ i}\right)$ given in eq.\ (1.4). Using it,
with $\langle\v r|\nu_i\rangle$ replaced by $\sum_{\v k}\langle\v r|\v
k\rangle\langle\v k|\nu_i\rangle$, it is possible to write this exchange
parameter in terms of the center-of-mass momenta and relative motion
indices of the ``in'' and ``out'' excitons, as
\begin{equation}
\lambda\left(^{\,\nu_n,\v K-\v p-\v Q\ \ \nu_j,\v K-\v p}_{\nu_m,\v K+\v
p+\v Q\ \ \nu_i,\v K+\v p}\right)=\sum_{\v k}\langle\nu_m|\v k+\frac{\v
P_-}{2}\rangle\,\langle\nu_n|\v k-\frac{\v P_-}{2}\rangle\,\langle\v
k+\frac{\v P_+}{2}|\nu_i\rangle\,\langle\v k-\frac{\v P_+}{2}|\nu_j\rangle\
,
\end{equation}
where $\v P_{\pm}=2\alpha_h\v p+(\alpha_h\pm\alpha_e)\v Q$. Note that this
exchange parameter does not depend on the total center-of-mass momentum
$2\v K$ of the ``in'' and ``out'' excitons, as physically expected.

The link between $\lambda\left(^{n\ j}_{m\ i}\right)$ and the
possibility to form excitons with different pairs also appears if we
couple the pairs of two excitons in different ways. Indeed, by using
eq.\ (1.1), we find
\begin{equation}
B_i^\dag B_j^\dag=\sum_{\v k_{e_i},\v k_{e_j},\v k_{h_i},\v k_{h_j}}
\langle\v k_{e_i},\v k_{h_i}|\phi_i\rangle\,\langle\v k_{e_j},\v k_{h_j}
|\phi_j\rangle\ a_{\v k_{e_i}}^\dag b_{\v k_{h_i}}^\dag a_{\v k_{e_j}}
^\dag b_{\v k_{h_j}}^\dag\ .
\end{equation}
If we now use the free pair $a_{\v k_{e_i}}^\dag b_{\v k_{h_j}}^\dag$
to form the $m$ exciton and the free pair $a_{\v k_{e_j}}^\dag b_{\v k
_{h_i}}^\dag$ to form the $n$ exciton, according to eq.\ (A.5) we get
\begin{eqnarray}
B_i^\dag B_j^\dag &=& -\sum_{mn}B_m^\dag B_n^\dag\sum_{\v k_{e_i},
\v k_{e_j},\v k_{h_i},\v k_{h_j}}\langle\v k_{e_i},\v k_{h_i}|\phi_i
\rangle\langle\v k_{e_j},\v k_{h_j}|\phi_j\rangle\langle\phi_m|
\v k_{e_i},\v k_{h_j}\rangle\langle\phi_n|\v k_{e_j},\v k_{h_i}
\rangle\nonumber
\\ &=& -\sum_{mn}\lambda\left(^{n\ j}_{m\ i}\right)
\,B_m^\dag B_n^\dag= -\sum_{mn}\lambda\left(^{m\ j}_{n\ i}\right)
\,B_m^\dag B_n^\dag\ ,
\end{eqnarray}
since $B_m^\dag B_n^\dag=B_n^\dag B_m^\dag$.

Let us end this part on the exchange parameter by deriving a quite
useful relation on a sum of $\lambda$'s. Starting from eq.\ (A.6)
and using the closure relation for excitons, namely $\sum_m\langle\v
p_e,\v p_h|
\phi_m\rangle
\langle\phi_m|\v k_e,\v k_h\rangle=\delta_{\v p_e,\v
k_e}\delta_{
\v p_h,\v k_h}$ and the one for free pairs, \linebreak$\sum_{\v k_e,\v
k_h}\langle\phi_p|\v k_e,\v k_h
\rangle\langle\v k_e, \v k_h|\phi_i\rangle=\delta_{pi}$, we immediately
get equation (1.5), which
shows that two hole exchanges reduce to an identity, as physically
reasonable. This however has, as a bad consequence, the fact that
counting the number of $\lambda$'s in a given quantity, does not amount
to count the number of exchanges between the excitons involved.

\section{Direct Coulomb scattering}

Let
$H=H_\mathrm{e}+H_\mathrm{h}+\mathcal{V}_\mathrm{ee}+\mathcal{V}
_\mathrm{hh}+\mathcal{V}_\mathrm{eh}$ be the semiconductor Hamiltonian. The
commutator of the exciton creation operator $B_i^\dag$ with the electron
kinetic part gives
\begin{equation}
\left[H_\mathrm{e},B_i^\dag\right] = \sum_{\v p,\v k_e, \v k_h}
\epsilon_{\v p}^{(e)}\,\langle\v k_e,\v k_h|\phi_i\rangle\,[a_{\v
p}^\dag a_{\v p}, a_{\v k_e}^\dag b_{\v k_h}^\dag]=\sum_{\v k_e,\v
k_h}\epsilon_{\v k_e} ^{(e)}\,\langle\v k_e,\v k_h|\phi_i\rangle\,a_{\v
k_e}^\dag b_{\v k_h} ^\dag\ ,
\end{equation}
with a similar result for $H_\mathrm{h}$. The electron-hole Coulomb
part gives
\begin{equation}
\left[\mathcal{V}_\mathrm{eh},B_i^\dag\right]=-\sum_{\v p_e,\v p_h,\v k_e,\v
k_h,
\v q}V_{\v q}\,\langle\v k_e,\v k_h|\phi_i\rangle\,[a_{\v p_e+\v q}
^\dag b_{\v p_h-\v q}^\dag b_{\v p_h} a_{\v p_e},a_{\v k_e}^\dag
b_{\v k_h}^\dag]\ ,
\end{equation}
the commutator between free pairs being equal to
\begin{eqnarray}
\left[a_{\v p_e+\v q}
^\dag b_{\v p_h-\v q}^\dag b_{\v p_h} a_{\v p_e},a_{\v k_e}^\dag
b_{\v k_h}^\dag\right]=\delta_{\v p_e,\v k_e}\delta_{\v p_h,\v k_h}\,
a_{\v k_e+\v q}^\dag b_{\v k_h-\v q}^\dag\hspace{5cm}\nonumber
\\ +\delta_{\v p_e,\v k_e}\,a_{\v k_e+\v q}^\dag b_{\v k_h}^\dag
b_{\v p_h-\v q}^\dag b_{\v p_h}+\delta_{\v p_h,\v k_h}\,a_{\v k_e}
^\dag b_{\v k_h-\v q}^\dag a_{\v p_e+\v q}^\dag a_{\v p_e}\ .
\end{eqnarray}
To go further, we can note that, for excitons eigenstates of
the semiconductor Hamiltonian, we have
\begin{equation}
H\,B_i^\dag|v\rangle=(H_\mathrm{e}+H_\mathrm{h}+\mathcal{V}_\mathrm{eh})
B_i^\dag|v\rangle=[(H_\mathrm{e}+H_\mathrm{h}
+\mathcal{V}_\mathrm{eh}),B_i^\dag]|v\rangle=E_i\,B_i^\dag|v\rangle\ ,
\end{equation}
with $E_i=\Delta+\varepsilon_{\nu_i}+Q_i^2/2(m_e+m_h)$, where $\Delta$ is the band gap ; so
that, if we insert eqs.\ (B.1-3) into this eq.\ (B.4), we find that the
$|\phi_i\rangle$'s are such that
\begin{equation}
(\epsilon_{\v k_e}^{(e)}+\epsilon_{\v k_h}^{(h)})\,\langle\v k_e,\v k_h
|\phi_i\rangle-\sum_{\v q}V_{\v q}\,\langle\v k_e-\v q,\v k_h +\v q|
\phi_i\rangle=E_i\,\langle\v k_e,\v k_h|\phi_i\rangle\ .
\end{equation}
Consequently, the commutator $[(H_\mathrm{e}+H_\mathrm{h}
+\mathcal{V}_\mathrm{eh}),B_i^\dag]|$ eventually reads
\begin{equation}
\left[(H_\mathrm{e}+H_\mathrm{h}+\mathcal{V}_\mathrm{eh}),B_i^\dag\right]=
E_i\,B_i^\dag+V_i^{\dag(eh)}\ ,
\end{equation}
with $V_i^{\dag(eh)}$ given by
\begin{equation}
V_i^{\dag(eh)}=-\sum_{\v q,\v k_e,\v k_h}V_{\v q}\,\langle\v k_e,\v k_h
|\phi_i\rangle\,\left(a_{\v k_e+\v q}^\dag b_{\v k_h}^\dag\sum_{\v p_h}
b_{\v p_h-\v q}^\dag b_{\v p_h}+a_{\v k_e}^\dag b_{\v k_h-\v q}^\dag
\sum_{\v p_e}a_{\v p_e+\v q}^\dag a_{\v p_e}\right)\ .
\end{equation}
In a similar way, we find
\begin{equation}
\left[\mathcal{V}_\mathrm{ee},B_i^\dag\right]=V_i^{\dag(ee)}=\sum_{\v q,\v
k_e,
\v k_h}V_{\v q}\,\langle\v k_e,\v k_h|\phi_i\rangle\,a_{\v k_e+\v q}
^\dag b_{\v k_h}^\dag\sum_{\v p_e}a_{\v p_e-\v q}^\dag a_{\v p_e}\ ,
\end{equation}
\begin{equation}
\left[\mathcal{V}_\mathrm{hh},B_i^\dag\right]=V_i^{\dag(hh)}=\sum_{\v q,\v
k_e,
\v k_h}V_{\v q}\,\langle\v k_e,\v k_h|\phi_i\rangle\,a_{\v k_e}
^\dag b_{\v k_h+\v q}^\dag\sum_{\v p_h}b_{\v p_h-\v q}^\dag b_{\v p_h}\
.
\end{equation}
So that we end with
\begin{equation}
\left[H,B_i^\dag\right]=E_i\,B_i^\dag+V_i^\dag\ ,
\end{equation}
\begin{equation}
V_i^\dag=V_i^{\dag(ee)}+V_i^{\dag(hh)}+V_i^{\dag(eh)}\ .
\end{equation}

We now turn to the commutator of this creation potential with the exciton
creation operator. From eqs.\ (B.8) and (1.1), we get
\begin{equation}
\left[V_i^{\dag(ee)},B_j^\dag\right]=\sum_{\v q,\v k_e,\v k_h,\v k'_e,
\v k'_h}V_{\v q}\,\langle\v k_e,\v k_h|\phi_i\rangle\langle\v k'_e,\v
k'_h|\phi_j\rangle\,a_{\v k_e+\v q}^\dag b_{\v k_h}^\dag
a_{\v k'_e-\v q}^\dag b_{\v k'_h}^\dag\ .
\end{equation}
We can rewrite the free pair operators $a_{\v k_e+\v
q}^\dag b_{\v k_h}^\dag$ and $a_{\v k'_e-\v q}^\dag b_{\v k'_h}^\dag$
in terms of creation operators for $(m,n)$ excitons,
according to eq.\ (A.5). This leads to
\begin{equation}
\left[V_i^{\dag(ee)},B_j^\dag\right]=\sum_{mn}\xi^{\mathrm{dir}}_{ee}
\left(_{m\ i}^{n\ j}\right)\,B_m^\dag B_n^\dag\ ,
\end{equation}
where the direct Coulomb scattering due to electron-electron
interaction is given by
\begin{equation}
\xi^{\mathrm{dir}}_{ee}\left(_{m\ i}^{n\ j}\right)=\sum_{\v q,\v
k_e,\v k_h,\v k'_e,\v k'_h}V_{\v q}\,\langle\v k_e,\v k_h|\phi_i\rangle
\langle\v k'_e,\v k'_h|\phi_j\rangle\langle\phi_m|\v k_e+\v q,\v k_h
\rangle\langle\phi_n|\v k'_e-\v q,\v k'_h\rangle\ .
\end{equation}
If we go to real space, this scattering reads
\begin{equation}
\xi^{\mathrm{dir}}_{ee}\left(_{m\ i}^{n\ j}\right)=\int d\v r_e\,
d\v r_{e'}\,d\v r_h\,d\v r_{h'}\,\phi_m^\ast(\v
r_e,\v r_h)\,\phi_n^\ast\v r_{e'},\v r_{h'})\,
\phi_i(\v r_e,\v r_h)\,\phi_j(\v r_{e'},\v r_{h'})\,V_{ee'}\ ,
\end{equation}
where $V_{ee'}=e^2/|\v r_e-\v r_{e'}|=\sum_{\v q}V_{\v q}\,
e^{i\v q.(\v r_e-\v r_{e'})}$ is the Coulomb interaction between the
electrons $e$ and $e'$. As for the exchange parameter, we can rewrite this
scattering in terms of the center-of-mass momenta and relative motion
indices as
\begin{eqnarray}
\xi^{\mathrm{dir}}_{ee}\left(^{\,\nu_n,\v K-\v p-\v Q\ \ \nu_j,\v K-\v p}
_{\nu_m,\v K+\v p+\v Q\ \ \nu_i,\v K+\v p}\right)&=&V_{\v Q}\sum_{\v k_i}
\langle\nu_m|\v k_i+\alpha_h\v Q\rangle\,\langle\v k_i|\nu_i\rangle\sum_{\v
k_j}\langle\nu_n|\v k_j-\alpha_h\v Q\rangle\,\langle\v k_j|\nu_j\rangle
\nonumber\\ &=& V_{\v Q}\,\langle\nu_m|e^{i\alpha_h\v Q.\v r}|\nu_i\rangle\,
\langle\nu_n|e^{-i\alpha_h\v Q.\v r}|\nu_j\rangle\ .
\end{eqnarray}

By using the same procedures for $V_i^{\dag (hh)}$ and $V_i^{\dag
(eh)}$, it is easy to recover eq.\ (1.8),
where, in real space, the direct Coulomb scattering between two excitons is given by eq.\
(1.9). In terms of the center-of-mass momenta and relative motion indices, this direct
Coulomb scattering appears as
\begin{equation}
\xi^{\mathrm{dir}}\left(^{\,\nu_n,\v K-\v p-\v Q\ \ \nu_j,\v K-\v p}
_{\nu_m,\v K+\v p+\v Q\ \ \nu_i,\v K+\v p}\right)=V_{\v Q}
\langle\nu_m|e^{i\alpha_h\v Q.\v r}-e^{-i\alpha_e\v Q.\v r}|\nu_i\rangle\,
\langle\nu_n|e^{-i\alpha_h\v Q.\v r}-e^{i\alpha_e\v Q.\v r}|\nu_j\rangle\ .
\end{equation}
Note that $\xi^\mathrm{dir}$ depends neither on the total momentum $2\v K$
of the ``in'' and ``out'' excitons, nor on the center-of-mass momenta of
the ``in'' excitons separately, namely on $\v p$, but just on the momentum
transfer $\v Q$.

We can note that, when one of the excitons stays unchanged, we have
\begin{equation}
\xi^\mathrm{dir}\left(_{i\ i}^{n\ j}\right)=0\ ,
\end{equation}
as can be seen by interchanging $e$ and $h$ in the
integral of eq.\ (1.9). Indeed, $\phi_i^\ast(\v r_e,
\v r_h)\phi_i(\v r_e,\v r_h)
=|\langle\v r_e-\v r_h|\varphi_{\nu_i}\rangle|^2$ stays unchanged
under this manipulation, whatever the parity of the relative motion
wave function of the $i$ exciton is. This result physically comes from the fact that,
in the case of excitons, the repulsion between identical fermions is as
large as the attraction between different fermions. Let us stress that
this property is no
more valid for ``cold atom'' composite bosons, which only have an
attractive part between different fermions in their interaction.

\section{Exchange Coulomb scatterings}

 From the
direct Coulomb scatterings $\xi^\mathrm{dir}\left(_{m\ i}^{n\
j}\right)$ and the exchange parameters $\lambda\left(_{m\ i}^{n\
j}\right)$, we can construct two rather important exchange scatterings
defined in eqs. (1.10) and (1.11), in which
the carrier exchange takes place after or before the Coulomb interaction.
By using eq.\ (1.5), it is easy to show that we also have
\begin{eqnarray}
\xi^\mathrm{dir}\left(_{m\ i}^{n\
j}\right)&=&\sum_{rs}\lambda\left(_{m\ r}^{n\
s}\right)\,\xi^\mathrm{in}\left(_{r\ i}^{s\
j}\right)\ ,\\ &=&
\sum_{rs}\xi
^\mathrm{out}\left(_{m\ r}^{n\
s}\right)\,\lambda\left(_{r\ i}^{s\
j}\right)\ .
\end{eqnarray}

 From the definitions of $\lambda$ and $\xi^\mathrm{dir}$ in $\v r$ space
given in eqs.\ (1.4) and (1.9) and the fact that $\sum_s\phi_s^\ast(\v
r_{e_1},\v r_{h_1})\phi_s(\v r_{e_2},\v r_{h_2})=\delta(\v r_{e_1}-\v
r_{e_2})\,\delta(\v r_{h_1}-\v r_{h_2})$, it is easy to recover the
expressions of $\xi^\mathrm{in}\left(_{m\ i}^{n\ j}\right)$ and
$\xi^\mathrm{out}\left(_{m\ i}^{n\ j}\right)$ given in eqs.\ (1.10,11).
These equations show that, while
\begin{equation}
\xi^\mathrm{out}\left(_{m\ i}^{n\ j}\right)=\left[\xi^\mathrm{in}
\left(_{i\ m}^{j\ n}\right)\right]^\ast\ ,
\end{equation}
we have for the contributions coming from $V_{ee'}$ and $V_{hh'}$
separately
\begin{equation}
\xi^\mathrm{out}_{cc'}\left(_{m\ i}^{n\ j}\right)=
\xi^\mathrm{in}_{cc'}\left(_{m\ i}^{n\ j}\right)\ ,
\end{equation}
with $c=e$ or $h$. This identity comes from the fact that the
electron-electron and hole-hole scatterings are between both, the ``in'' and
the ``out'' excitons, whatever the position of the carrier exchange is,
while this is not true for the electron-hole parts.
In terms of the center-of-mass momenta and relative motion indices, these
$\xi^\mathrm{in}_{cc'}$'s appear as
\begin{eqnarray}
\xi^\mathrm{in}_{cc'}\left(^{\,\nu_n,\v K-\v p-\v Q\ \ \nu_j,\v K-\v
p}_{\nu_m,\v K+\v p+\v Q\ \ \nu_i,\v K+\v p}\right)=\sum_{\v
k}\sum_{\v q\neq\v 0}V_{\v q}
\langle\nu_m|\v k+\frac{\v P_-+\v q}{2}\rangle\,\langle\nu_n|\v k-\frac{\v
P_-+\v q}{2}\rangle\nonumber\\ \times\
\langle\v k+\frac{\v P_+\mp
\v q}{2}|\nu_i\rangle\,\langle\v k-\frac{\v P_+\mp \v q}{2}|\nu_j\rangle\ ,
\end{eqnarray}
with the upper sign for $(ee')$ and the lower sign for $(hh')$, the momenta
$\v P_{\pm}$ being defined as for the exchange parameter $\lambda$ (see
eq.\ (A.7)).

Similarly, the contributions to $\xi^\mathrm{in}$ coming from Coulomb
interactions between electron and hole read
\begin{eqnarray}
\xi^\mathrm{in}_{cd'}\left(^{\,\nu_n,\v K-\v p-\v Q\ \ \nu_j,\v K-\v
p}_{\nu_m,\v K+\v p+\v Q\ \ \nu_i,\v K+\v p}\right)=-\sum_{\v
k}\sum_{\v q\neq\v 0}V_{\v q}
\langle\nu_m|\v k+\frac{\v P_-+\v q}{2}\rangle\,\langle\nu_n|\v k-\frac{\v
P_-+\v q}{2}\rangle\nonumber\\ \times\
\langle\v k+\frac{\v P_+\mp
\v q}{2}|\nu_i\rangle\,\langle\v k-\frac{\v P_+\pm \v q}{2}|\nu_j\rangle\ ,
\end{eqnarray}
with the upper sign for $(eh')$ and the lower sign for $(he')$.

It is of interest to note that, in these $\xi^\mathrm{in}_{cd'}$, the
sum over $\v q$ can be readily done through
\begin{equation}
\epsilon_{\v k}\,\langle\v k|\nu\rangle-\sum_{\v q}V_{\v q}\,\langle\v k+\v
q|\nu\rangle=\varepsilon_\nu\,\langle\v k|\nu\rangle\ ,
\end{equation}
with $\epsilon_{\v k}=k^2/2\mu_X$ and $\mu_X^{-1}=m_e^{-1}+m_h^{-1}$, which
follows from eq.\ (B.5). By setting $\v k'=\v k-\v q/2$, we then find
\begin{eqnarray}
\xi^\mathrm{in}_{eh'}\left(^{\,\nu_n,\v K-\v p-\v Q\ \ \nu_j,\v K-\v
p}_{\nu_m,\v K+\v p+\v Q\ \ \nu_i,\v K+\v p}\right)=\sum_{\v k'}
(\varepsilon_{\nu_m}-\epsilon_{\v k'+\v P_-/2})\,\langle\nu_m|\v k'
+\frac{\v P_-}{2}\rangle\,
\langle\nu_n|\v k'-\frac{\v P_-}{2}\rangle\nonumber\\ \times\
\langle\v k'+\frac{\v P_+}{2} |\nu_i\rangle\,
\langle\v k'-\frac{\v P_+}{2}|\nu_j\rangle\ .
\end{eqnarray}
In the same way, by setting $\v k'=\v k+\v q/2$, we find
\begin{eqnarray}
\xi^\mathrm{in}_{he'}\left(^{\,\nu_n,\v K-\v p-\v Q\ \ \nu_j,\v K-\v
p}_{\nu_m,\v K+\v p+\v Q\ \ \nu_i,\v K+\v p}\right)=\sum_{\v k'}
(\varepsilon_{\nu_n}-\epsilon_{\v k'-\v P_-/2})\,\langle\nu_m|\v k'
+\frac{\v P_-}{2}\rangle\,
\langle\nu_n|\v k'-\frac{\v P_-}{2}\rangle\nonumber\\ \times\
\langle\v k'+\frac{\v P_+}{2} |\nu_i\rangle\,
\langle\v k'-\frac{\v P_+}{2}|\nu_j\rangle\ .
\end{eqnarray}
Note that these eqs.\ (C.8) and (C.9) are similar to the expression
(A.7) of the exchange parameter, except for the prefactors.

\section{Energy-like Pauli scattering}

With $\lambda\left(_{m\ i}^{n\ j}\right)$ alone, we can construct an
energy-like scattering defined in eq.\ (1.12), which does not contain any
Coulomb scattering between excitons explicitly. However, this
$\mathcal{E}\left(_{m\ i}^{n\ j}\right)$ scattering which seems to
only rely on the composite boson character of the excitons through $\lambda\left(_{m\ i}^{n\
j}\right)$, is nothing but the difference between the two exchange Coulomb scatterings, as written
in eq.\ (1.13). In order to derive this relation, we write
$\xi^\mathrm{in}\left(_{m\ i}^{n\ j}\right)-\xi^\mathrm{out}\left(_{m\
i}^{n\ j}\right)$, using eqs.\ (1.10,11). This leads to
\begin{eqnarray}
\xi^\mathrm{in}
\left(_{m\ i}^{n\ j}\right)-\xi^\mathrm{out}\left(_{m\ i}^{n\
j}\right)=\int d\v r_e\,d\v r_{e'}\,d\v r_h\,d\v r_{h'}\,
\phi_m^\ast(\v r_e,\v r_{h'})\,\phi_n^\ast(\v r_{e'},\v r_h)\hspace{2cm}
\nonumber
\\ \times\left[V_{eh}+V_{e'h'}-V_{eh'}-V_{e'h}\right]\,
\phi_i(\v r_e,\v r_h)\,\phi_j(\v r_{e'},\v r_{h'})\ .
\end{eqnarray}
So that this difference only comes from electron-hole interactions, in
agreement with the fact that the contribution from electron-electron or
hole-hole interactions are similar for ``in'' and ``out'' Coulomb exchange
scatterings (see eq.\ (C.4)).

By turning to $\v k$ space, we find that the $V_{eh}$ term of the
above equation reads
\begin{eqnarray}
\int d\v r_e\,d\v r_{e'}\,d\v r_h\,d\v r_{h'}\,
\phi_m^\ast(\v r_e,\v r_{h'})\,\phi_n^\ast(\v r_{e'},\v r_h)\,V_{eh}\,
\phi_i(\v r_e,\v r_h)\,\phi_j(\v r_{e'},\v r_{h'})\hspace{4cm}\nonumber
\\ =\sum_{\v k_e,\v k_h,\v k'_e, \v k'_h,\v q}V_{\v q}
\langle\phi_m|\v k_e,\v k'_h\rangle\langle\phi_n|\v k'_e,\v k_h\rangle
\langle\v k_e-\v q,\v k_h+\v q|\phi_i\rangle
\langle\v k'_e,\v k'_h|\phi_j\rangle\hspace{1.3cm}\nonumber
\\ =\sum_{\v k_e,\v k_h,\v k'_e, \v k'_h}(\epsilon_{\v k_e}^{(e)}
+\epsilon_{\v k_h}^{(h)}-E_i)
\langle\phi_m|\v k_e,\v k'_h\rangle\langle\phi_n|\v k'_e,\v k_h\rangle
\langle\v k_e,\v k_h|\phi_i\rangle
\langle\v k'_e,\v k'_h|\phi_j\rangle\ ,\hspace{0.5cm}
\end{eqnarray}
where we have used eq.\ (B.5). By calculating the $V_{e'h'}$ (resp.\
$V_{eh'}$ and $V_{e'h}$) term of eq.\ (D.1) in a similar way, we find
that it reads as the last line of eq.\ (D.2) with the prefactor
$(\epsilon_{\v k_e}^{(e)} +\epsilon_{\v k_h}^{(h)}-E_i)$ replaced by
$(\epsilon_{\v k'_e}^{(e)} +\epsilon_{\v k'_h}^{(h)}-E_j)$ (resp.\
$(\epsilon_{\v k_e}^{(e)} +\epsilon_{\v k'_h}^{(h)}-E_m)$ and
$(\epsilon_{\v k'_e}^{(e)} +\epsilon_{\v k_h}^{(h)}-E_n)$). By adding
the four terms, we eventually get
\begin{eqnarray}
\xi^\mathrm{in}
\left(_{m\ i}^{n\ j}\right)-\xi^\mathrm{out}\left(_{m\ i}^{n\
j}\right) &=& (E_m+E_n-E_i-E_j)\nonumber
\\ &\times& \sum_{\v k_e,\v k_h,\v k'_e, \v k'_h}
\langle\phi_m|\v k_e,\v k'_h\rangle\langle\phi_n|\v k'_e,\v k_h\rangle
\langle\v k_e,\v k_h|\phi_i\rangle
\langle\v k'_e,\v k'_h|\phi_j\rangle\nonumber
\\ &=& (E_m+E_n-E_i-E_j)\,\lambda\left(_{m\
i}^{n\ j}\right)=\mathcal{E}\left(_{m\
i}^{n\ j}\right)\ .
\end{eqnarray}
Note that it is possible to recover this result directly from eqs.\ (C.3,4)
and (C.8,9).

As a useful consequence of this eq.\ (D.3), the ``in'' and ``out'' exchange
scatterings are equal if the energies of the ``in'' and ``out'' excitons
are equal. This in particular shows that they are equal for diagonal
processes:
\begin{equation}
\xi^\mathrm{in}\left(_{i\
i}^{j\ j}\right)=\xi^\mathrm{out}\left(_{i\
i}^{j\ j}\right)\ .
\end{equation}

\section{Key equations to get correlation effects and time evolution of
composite excitons}

Correlation effects between composite excitons are obtained from the
iteration of eq.\ (4.15).
This equation follows from the commutator $[H,B_i^\dag]$. Indeed, eq.\ (1.7)
gives
\begin{equation}
H\,B_i^\dag=B_i^\dag(H+E_i)+V_i^\dag\ ,
\end{equation}
so that
\begin{equation}
B_i^\dag(a-H-E_i)=(a-H)B_i^\dag+V_i^\dag\ .
\end{equation}
If we now multiply this equation by $(a-H)^{-1}$ on the left and
$(a-H-E_i)^{-1}$ on the right, we readily get eq.\ (4.15).

Equation (4.15) can be used to obtain the time evolution of exciton
states as an expansion in Coulomb scatterings. For that, we first note that
\begin{equation}
e^{-iHt}=-\int_{-\infty}^{+\infty}\frac{dx}{2i\pi}\,\frac{e^{-ixt}}
{x-H+i\eta}\ ,
\end{equation}
which is valid for any $t$ and $\eta$ $> 0$. This is easy to check either
by performing, in a formal way, the integration over the path made of the
real axis $(-\infty,+\infty)$ and the lower infinite half circle, or by
performing the same integration after having projected the operator at hand
over a complete basis made of the $H$ eigenstates. This path goes around the
pole $z=H-i\eta$, while it gives a negligible contribution over the circle
$z=R\,e^{i\theta}$ for $-\pi<\theta<0$ and $R\rightarrow +\infty$, since
$t>0$.

This leads to
\begin{eqnarray}
e^{-iHt}\,B_i^\dag =-\int_{-\infty}^{+\infty}\frac{dx}{2i\pi}
\,\frac{e^{-ixt}}{x-H+i\eta}\ B_i^\dag
= -\int_{-\infty}^{+\infty}\frac{dx}{2i\pi}\,e^{-ixt}\left[
B_i^\dag\,\frac{1}{x-H-E_i+i\eta}\right.\nonumber\\
+\left.\frac{1}{x-H+i\eta}\,V_i^\dag\,
\frac{1}{x-H-E_i+i\eta}\right]\ ,
\end{eqnarray}
due to eq.\ (4.15). From it, we readily recover eqs.\ (4.16,17), where the
operator
$W_i^\dag(t)$ gives zero when acting on vacuum.

\newpage

\begin{figure}[h]
\centerline{\scalebox{0.7}{\includegraphics{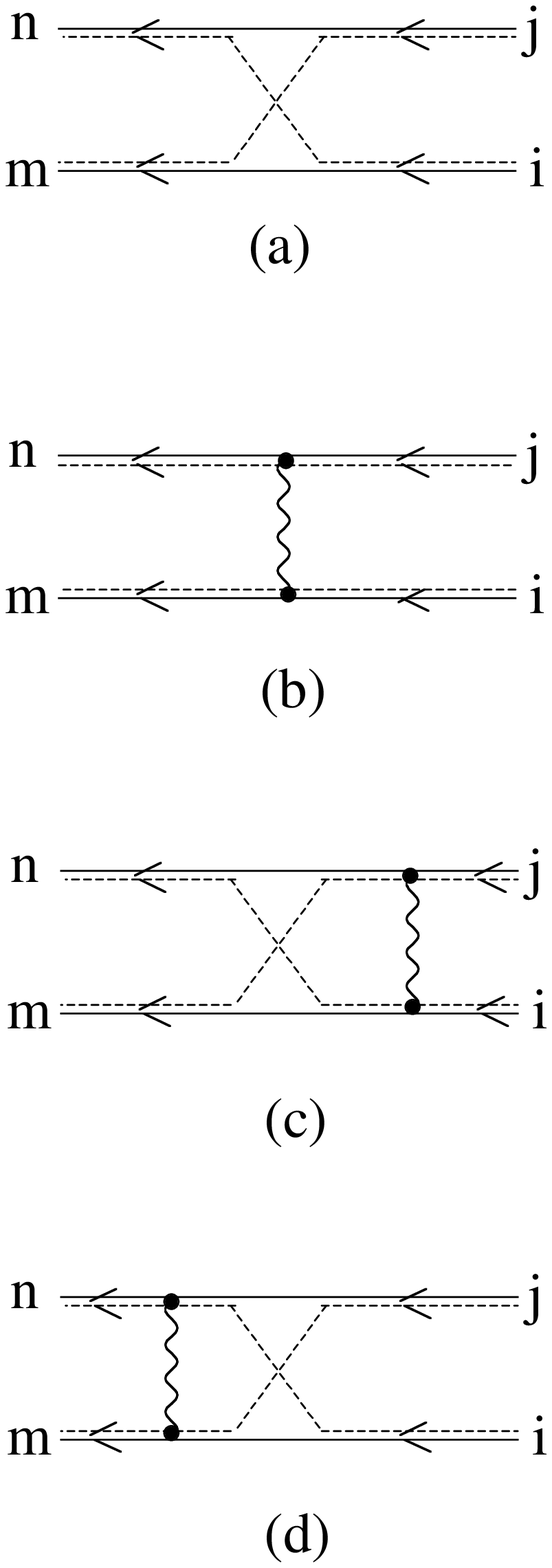}}}
\caption{(a): Pauli -- or exchange -- parameter $\lambda\left(^{n\ \,j}_{m\ i}\right)$ between two
``in'' excitons
$(i,j)$ giving rise to two ``out'' excitons $(m,n)$, as defined in eq.\ (1.4). (b): Direct Coulomb
scattering $\xi^\mathrm{dir}\left(^{n\ \,j}_{m\ i}\right)$ as defined in eq.\ (1.9). (c) and (d):
Coulomb exchange scatterings $\xi^\mathrm{in}\left(^{n\ \,j}_{m\ i}\right)$ and
$\xi^\mathrm{out}\left(^{n\ \,j}_{m\ i}\right)$ as defined in eqs.\ (1.10) and (1.11). In
$\xi^\mathrm{in}$, the Coulomb processes take place between the ``in'' excitons, while in
$\xi^\mathrm{out}$, they take place between the ``out'' excitons. In all these diagrams,
the solid lines represent the electrons while the dashed lines represent the holes, the wavy lines
representing the Coulomb interactions between the fermions of the excitons represented by double
electron-hole lines.}
\end{figure}

\clearpage

\begin{figure}[h]
\centerline{\scalebox{0.7}{\includegraphics{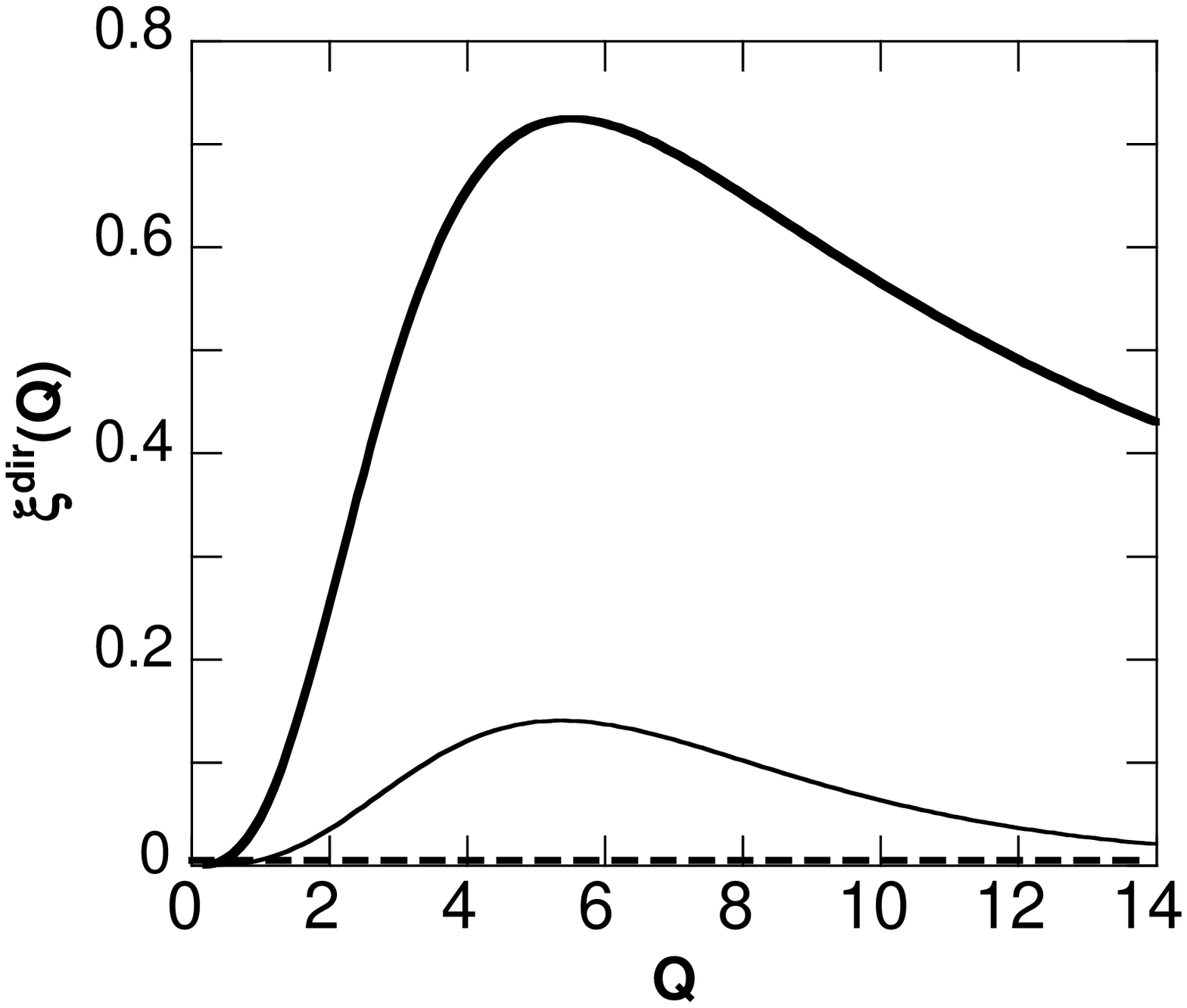}}}
\caption{Direct Coulomb scattering $\xi^\mathrm{dir}\left(^{\nu_0,-\v Q\ \ \nu_0,\v 0}_{\,\,
\nu_0,\v Q\ \ \
\nu_0,\v 0}\right)$ given in eqs.\ (6.1) and (6.3) for two quantum well ground states excitons with
same momentum taken as 0, scattered into two ground state excitons with momenta $(+\v Q)$ and $(-\v
Q)$ for three values of $m_h/m_e$. Thick solid line: $m_h\gg m_e$; thin solid line: $m_h=2m_e$;
thick dashed line: $m_h=m_e$. This plot, as all the other following plots, are done in reduced
units, namely, $(e^2/a_X)(a_X/L)^2$ for the scatterings and $a_X^{-1}$ for the momenta (see eq.\
6.3)), $a_X$ being the 3D exciton Bohr radius.}
\end{figure}

\clearpage

\begin{figure}[h]
\centerline{\scalebox{0.7}{\includegraphics{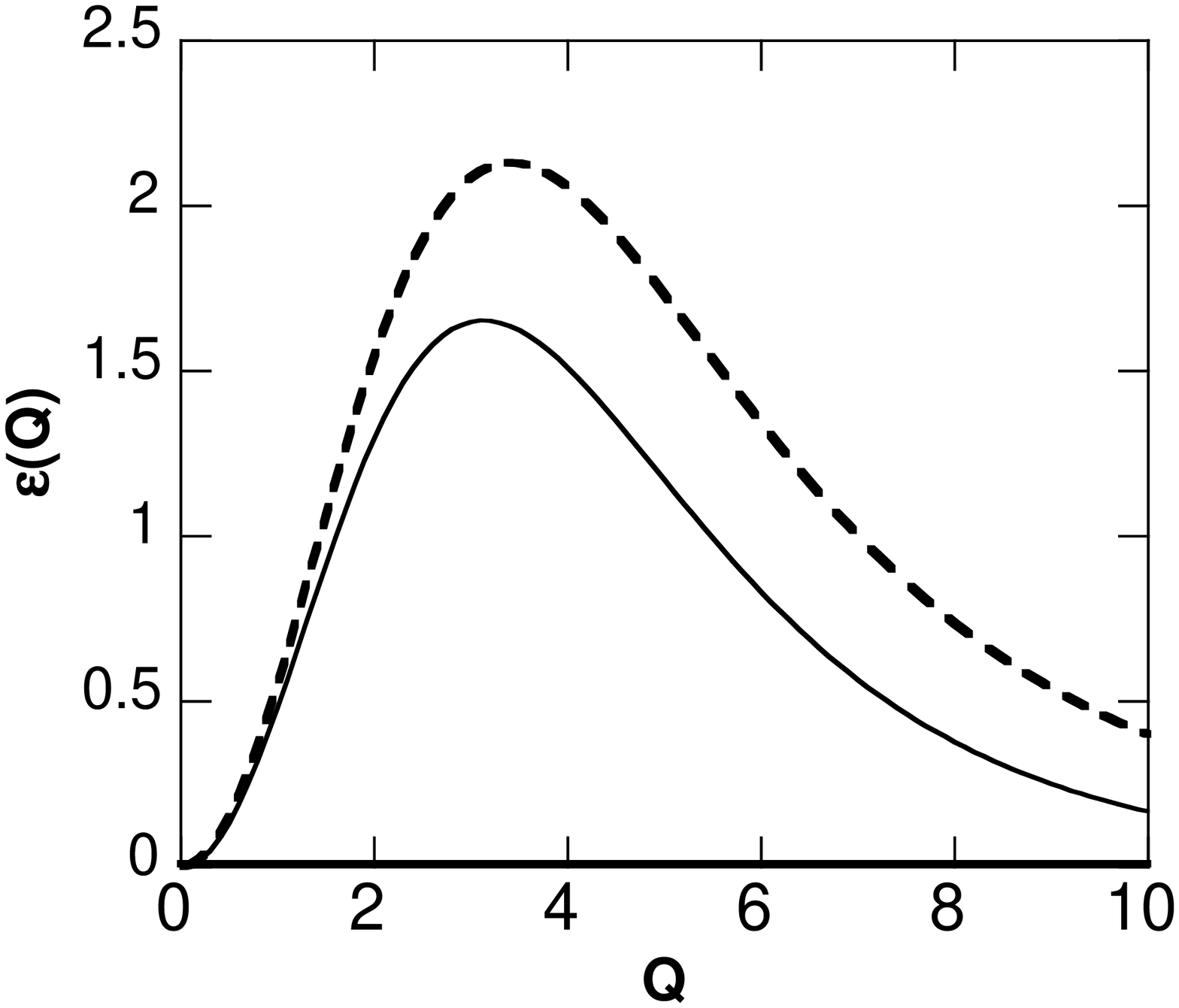}}}
\caption{Energy-like scattering $\mathcal{E}\left(^{\nu_0,-\v Q\ \ \nu_0,\v 0}_{\,\,
\nu_0,\v Q\ \ \ \nu_0,\v 0}\right)$, given in eqs.\ (6.4) and (6.6), constructed on the
dimensionless exchange parameter $\lambda\left(^{n\ \,j}_{m\ i}\right)$ (see eq.\ (1.12)), for
$m_h=m_e$, $m_h=2m_e$ and $m_h\gg m_e$ (same notations as in Fig.2).}
\end{figure}

\clearpage

\begin{figure}[h]
\centerline{\scalebox{0.7}{\includegraphics{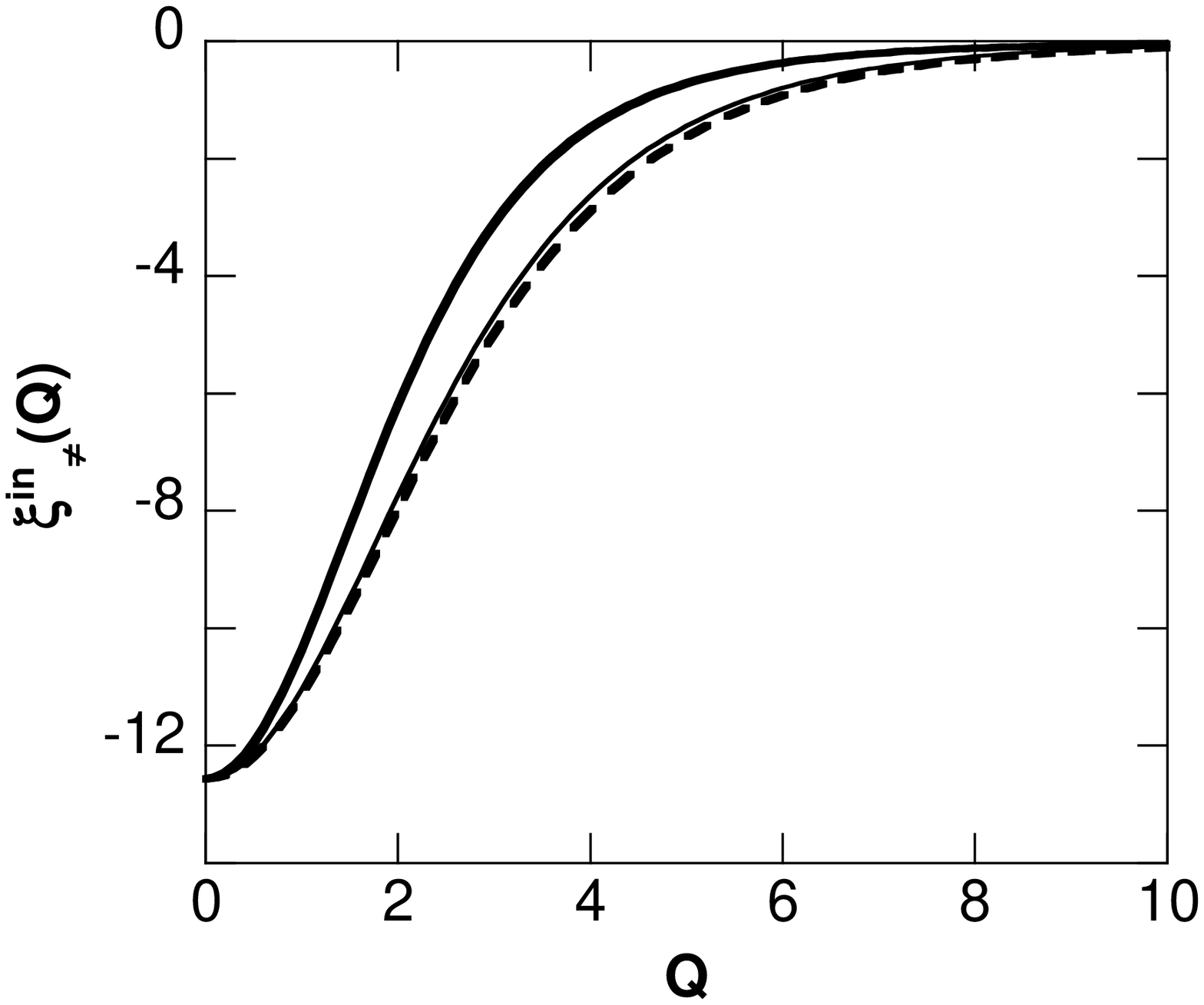}}}
\caption{Contribution to the ``in'' Coulomb exchange scattering $\xi^\mathrm{in}_{\neq}
\left(^{\nu_0,-\v Q\ \ \nu_0,\v 0}_{\,\,\nu_0,\v Q\ \ \ \nu_0,\v 0}\right)$, coming from
electron-hole interactions (see eqs.\ (6.7-8)), for $m_h\gg m_e$, $m_h=2m_e$ and
$m_h=m_e$ (same notations as in Fig.2).}
\end{figure}

\clearpage

\begin{figure}[h]
\centerline{\scalebox{0.7}{\includegraphics{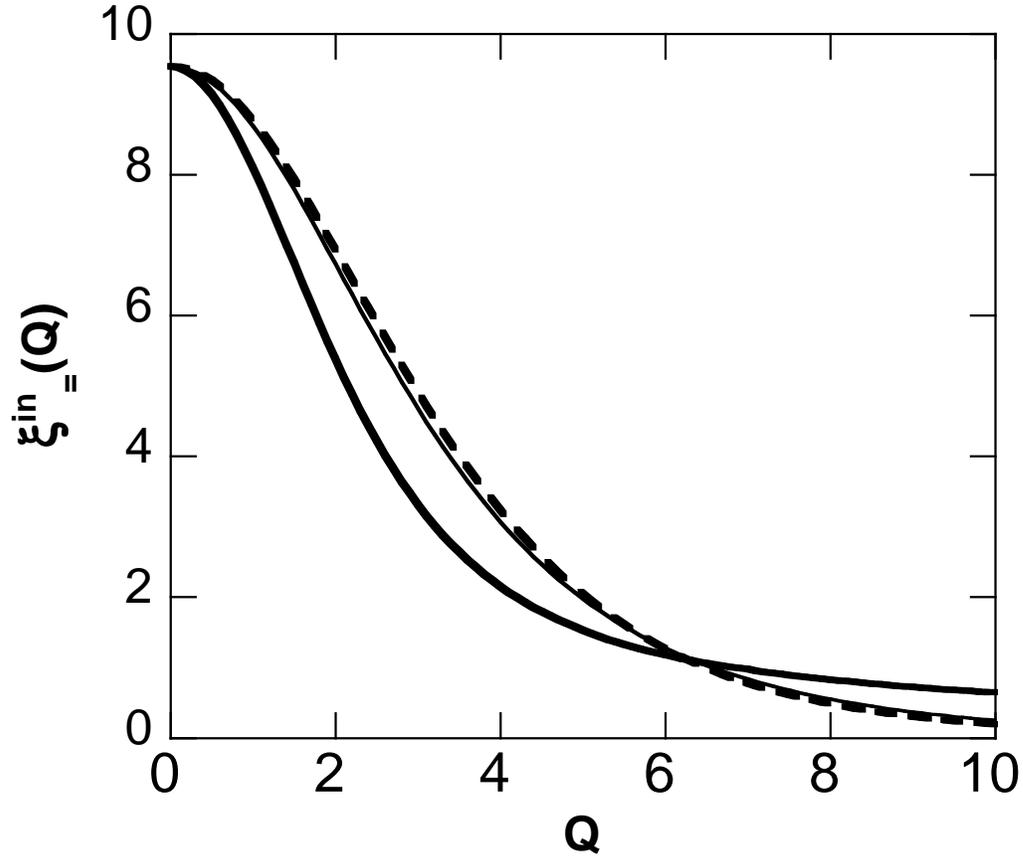}}}
\caption{Contribution to the ``in'' Coulomb exchange scattering $\xi^\mathrm{in}_{=}
\left(^{\nu_0,-\v Q\ \ \nu_0,\v 0}_{\,\,\nu_0,\v Q\ \ \ \nu_0,\v 0}\right)$, coming from
electron-electron and hole-hole interactions (see eqs.\ (6.9-10)), 
for $m_h\gg m_e$, $m_h=2m_e$ and $m_h=m_e$ (same notations as in Fig.2).}
\end{figure}

\clearpage
 
\begin{figure}[h]
\centerline{\scalebox{0.7}{\includegraphics{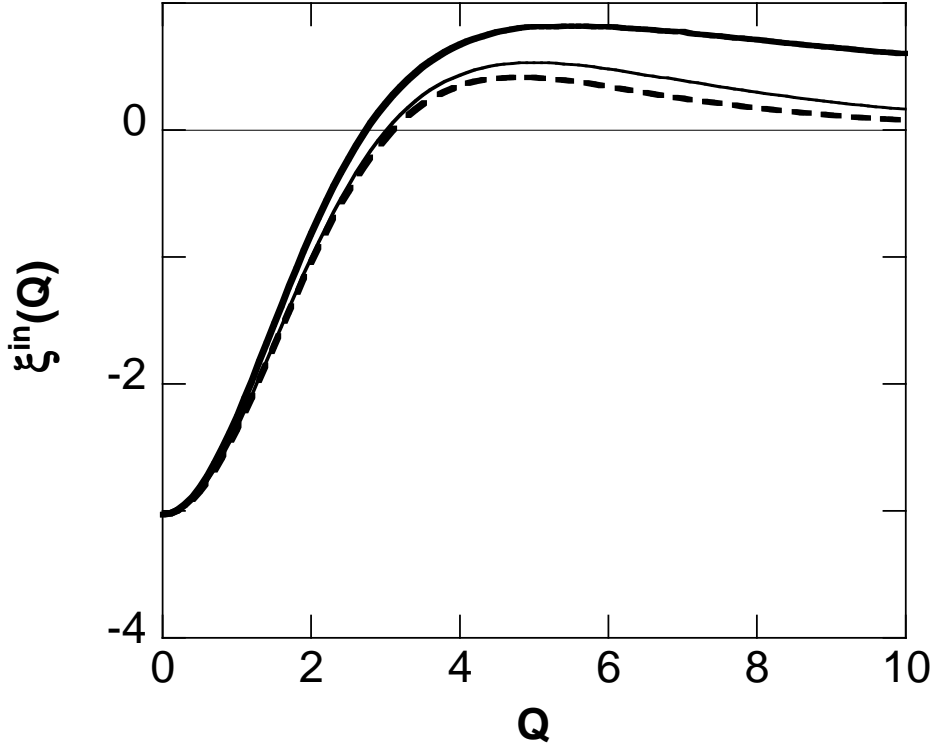}}}
\caption{``In'' Coulomb exchange scattering $\xi^\mathrm{in}
\left(^{\nu_0,-\v Q\ \ \nu_0,\v 0}_{\,\,\nu_0,\v Q\ \ \ \nu_0,\v 0}\right)$, resulting from all
Coulomb interactions between the ``in'' excitons (see eq.\ (6.11)), for $m_h\gg m_e$, $m_h=2m_e$
and $m_h=m_e$ (same notations as in Fig.2). Note that these scatterings cancel for a finite
momentum transfer $\v Q$.}
\end{figure}

\clearpage

\begin{figure}[h]
\centerline{\scalebox{0.7}{\includegraphics{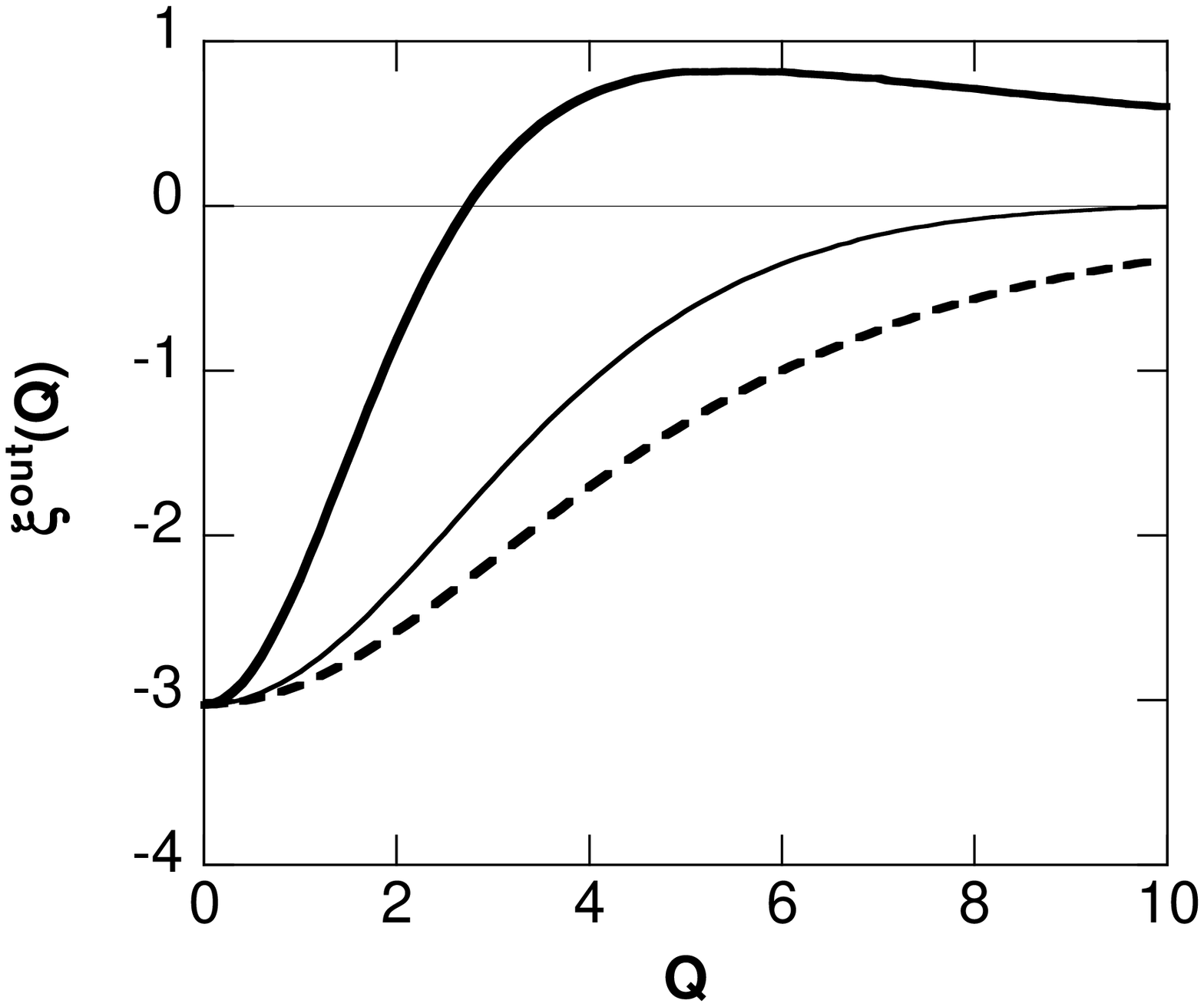}}}
\caption{``Out'' Coulomb exchange scattering $\xi^\mathrm{out}
\left(^{\nu_0,-\v Q\ \ \nu_0,\v 0}_{\,\,\nu_0,\v Q\ \ \ \nu_0,\v 0}\right)$, resulting from all
Coulomb interactions between the ``out'' excitons, for $m_h\gg m_e$, $m_h=2m_e$
and $m_h=m_e$ (same notations as in Fig.2). We see from Figs.6 and 7 that the ``in'' and ``out''
Coulomb exchange scatterings have similar behaviors while their precise values are different
except for
$m_h\gg m_e$ (thick solid line) or for zero momentum transfer $\v Q$.}
\end{figure}

\clearpage

\begin{figure}[h]
\centerline{\scalebox{0.7}{\includegraphics{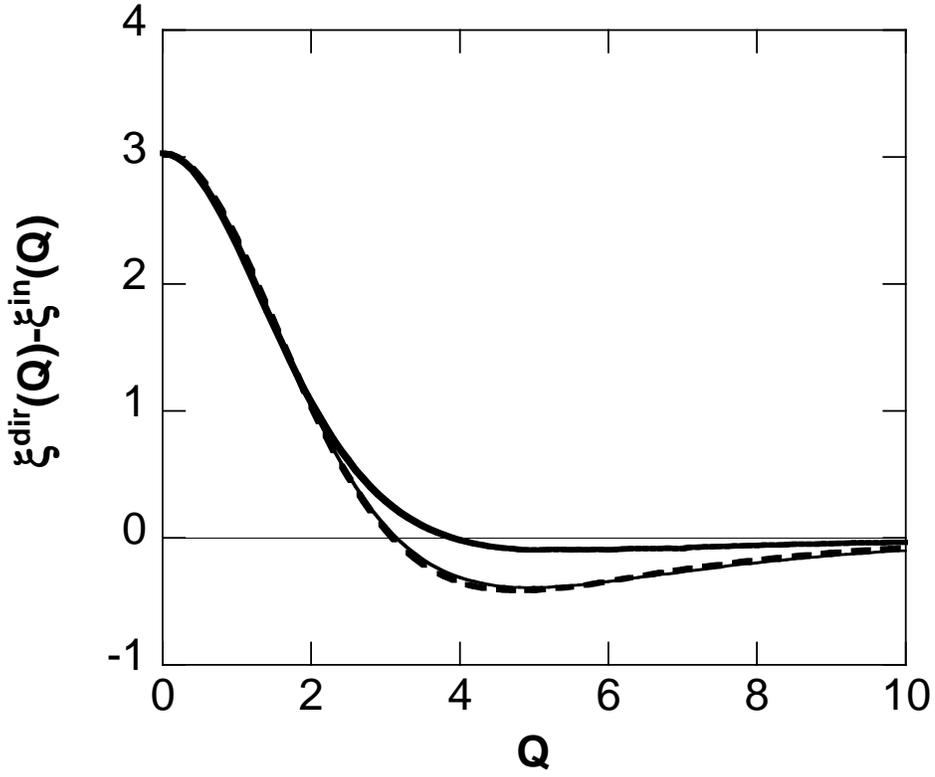}}}
\caption{Effective scattering $\xi^\mathrm{dir}-\xi^\mathrm{in}$ appearing in the transition
rate of two excitons with same initial momentum, as obtained from the many-body theory for
composite excitons, for
$m_h\gg m_e$,
$m_h=2m_e$ and $m_h=m_e$ (same notations as in Fig.2). We see that this effective scattering
cancels for a finite value of the momentum transfer $\v Q$, this value depending on the mass ratio.
}
\end{figure}

\clearpage

\begin{figure}[h]
\centerline{\scalebox{0.7}{\includegraphics{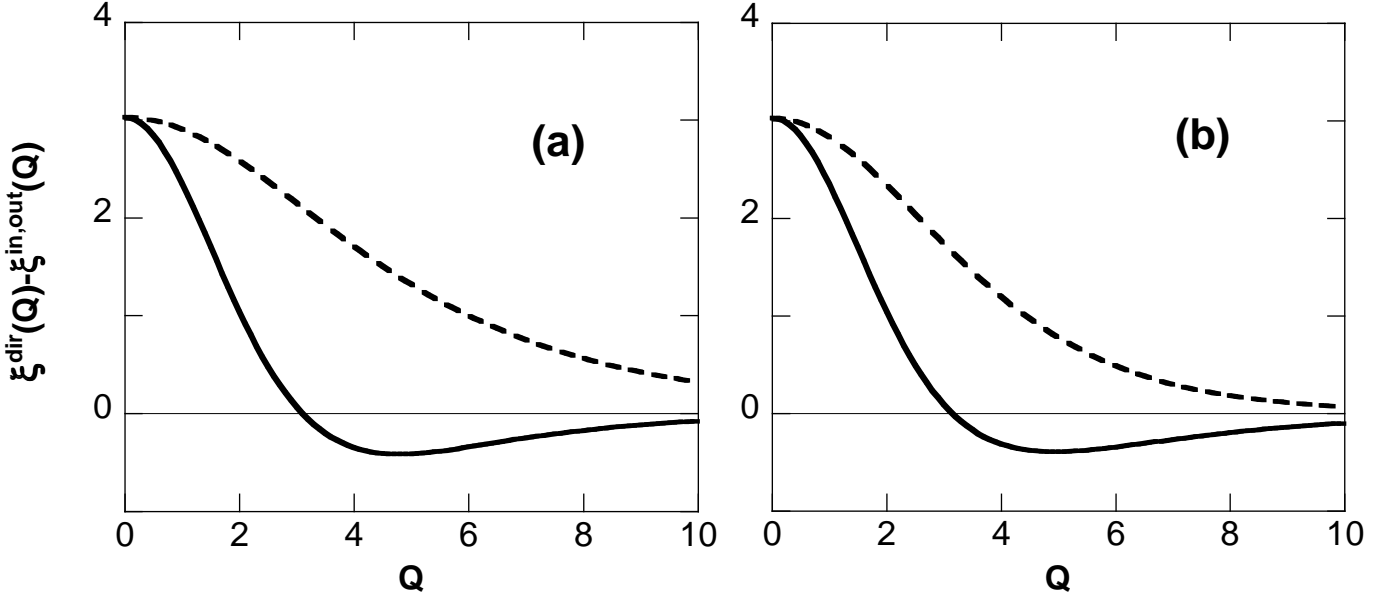}}}
\caption{Solid line : effective scattering $\xi^\mathrm{dir}-\xi^\mathrm{in}$, which appears in
the exact calculation of the transition rates of two excitons having same initial momentum, as a
function of the momentum transfer $\v Q$. Dashed line : effective scattering
$\xi^\mathrm{dir}-\xi^\mathrm{out}$, as obtained
from the effective bosonic Hamiltonian up to now used.
Fig.(a) corresponds to
$m_h=m_e$, while Fig.(b) corresponds to $m_h=2m_e$, the two effective scatterings being equal for
$m_h\gg m_e$, as seen from eq.\ (6.12). The noticeable discrepancy between these effective
scatterings questions the significance of the ``very good agreement with experiments'' obtained by
using such effective bosonic scatterings, since in real experimental conditions, the exciton 
mass ratio $m_h/m_e$ is large but finite.}
\end{figure}

\end{document}